\documentclass[preprint,12pt]{elsarticle}

\usepackage[utf8]{inputenc}

\usepackage{amssymb}
\usepackage{amsmath}
\usepackage[T1]{fontenc}

\usepackage[version=4]{mhchem}
\usepackage{enumitem,cleveref}
\usepackage{siunitx}
\DeclareSIUnit\angstrom{\text {Å}}
\usepackage{tabularx}
\usepackage{xcolor}
\graphicspath{{figures/}}
\RequirePackage{tikz}
\usetikzlibrary{arrows,positioning,calc,shapes.geometric}

\tikzstyle{labeled}=[execute at begin node=$\scriptstyle,
   execute at end node=$]
\RequirePackage{graphicx}

\journal{Computational Materials Science}

\begin{document}

\begin{frontmatter}



\title{Phonon density of states of magnetite (\ce{Fe3O4}) nanoparticles via molecular dynamics simulations.}


\author[ANSTO]{Pablo Galaviz} 
\author[UofW1,UofW2]{Kyle A. Portwin} 
\author[ANSTO]{Dehong Yu}
\author[ANSTO,UofW2]{Kirrily C. Rule}
\author[ANSTO,UofW2]{David L. Cortie}
\author[UofW1]{Zhenxiang Cheng}

\affiliation[ANSTO]{organization={Australian Centre for Neutron Scattering, Australian Nuclear Science and Technology Organisation},
            addressline={New Illawarra Road}, 
            city={Lucas Heights},
            postcode={2234}, 
            state={New South Wales},
            country={Australia}}
\affiliation[UofW1]{organization={Institute for Superconducting and Electronic Materials, Faculty of Engineering and Information Science, University of Wollongong},
            addressline={Innovation Campus, Squires Way}, 
            city={North Wollongong},
            postcode={2500}, 
            state={New South Wales},
            country={Australia}}
\affiliation[UofW2]{organization={School of Physics, Faculty of Engineering and Information Science, University of Wollongong},
            addressline={Northfields Avenue}, 
            city={Wollongong},
            postcode={2522}, 
            state={New South Wales},
            country={Australia}}


\begin{abstract}
This study presents a comprehensive computational investigation of magnetite nanoparticles, systematically evaluating a range of force fields against experimental results. We analyze the influence of particle size, temperature, and surface-adsorbed water molecules on the structural and dynamic properties of the nanoparticles. 
We performed classical molecular dynamics of nanoparticles and bulk magnetite and utilized density functional theory calculations for bulk magnetite for comparison.
Our results reveal that nanoparticle size and the presence of adsorbed water molecules have a pronounced impact on the density of states. Specifically, as the nanoparticle size is decreased, phonon modes exhibit significant broadening and softening, which is attributable to reduced phonon lifetimes resulting from enhanced boundary scattering. The incorporation of water further broadens the density of states and extends the spectra to higher energy regions. Temperature variations result in a slight broadening and softening of the phonon density of states, particularly in the oxygen-dominated region, which is attributed to phonon anharmonicity. 

\end{abstract}

\begin{graphicalabstract}
\centering
\includegraphics[width=0.85\textwidth]{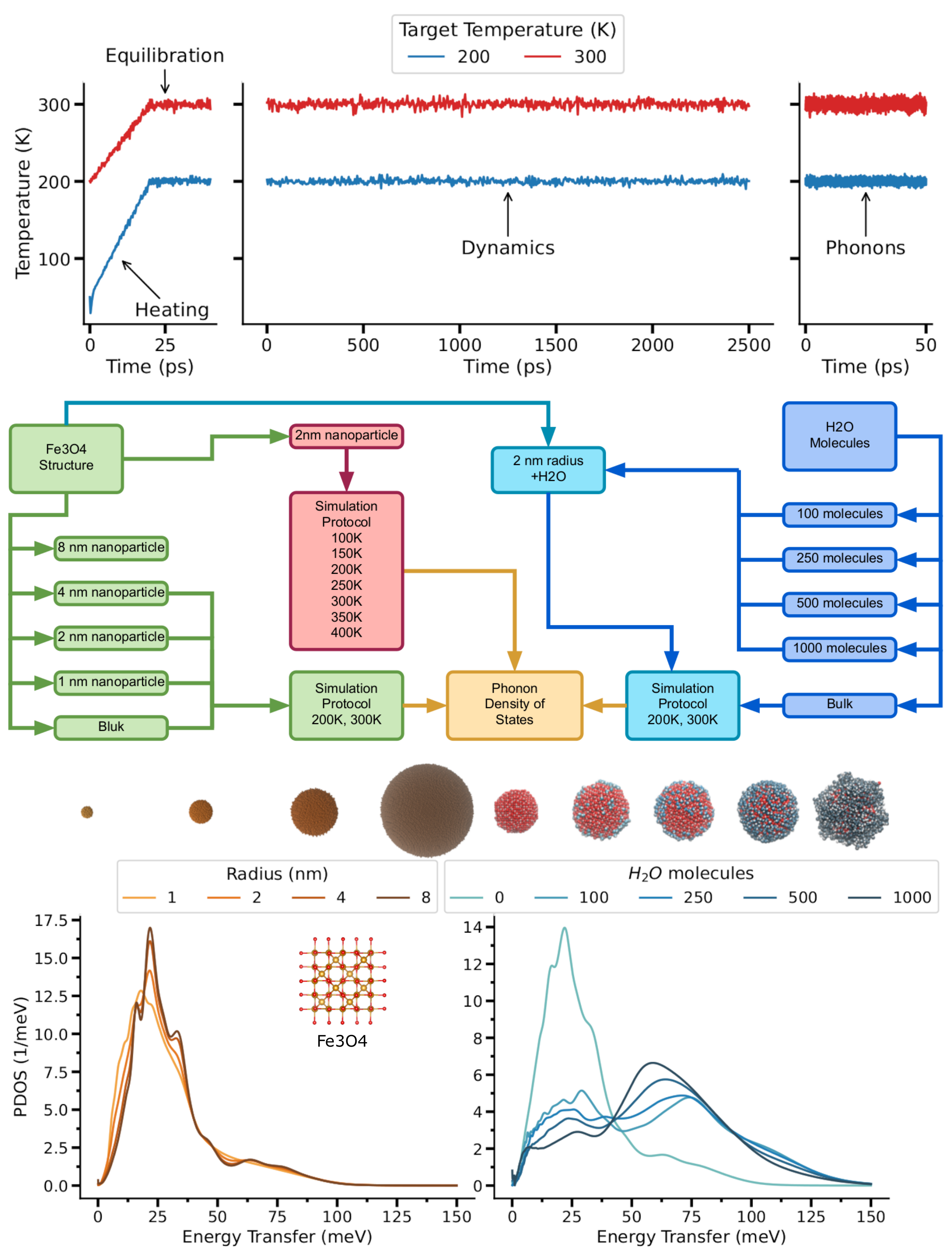}
\end{graphicalabstract}


\begin{highlights}
\item A range of classical force fields were tested to identify the optimum method to reproduce the phonon density of states that agrees relatively well with x-ray, neutron scattering, and density functional theory calculations.
\item The size of the magnetite nanoparticle significantly affects the phonon density of states, showing distinctive surface modes. 
\item The phonon density of states weakly depends on temperature between \qty{100}{K} and \qty{400}{K}.
\item Simulating an isolated particle or a cluster of particles generates a similar phonon density of states. 
\item The inclusion of a few surface \ce{H2O} molecules significantly affects the total density of states. Additionally, it creates a stronger dependency on temperature fluctuations. 
\end{highlights}


\end{frontmatter}



\section{Introduction}
\label{sec: introduction}

Nanoparticles are a broad class of materials with a size of less than \qty{100}{nm} that often exhibit unique physical and chemical properties that differ from a bulk material of the same composition. Nanoparticles have a large surface-to-volume ratio, which affects their optical, mechanical, magnetic, thermal, and electrical properties as a function of size. Nanoparticles have applications in medicine, electronics, energy harvesting, catalysis, and materials engineering \cite{KhaSaeKha2019, Alt2023}. We can define three regions in a nanoparticle: the core, which has the structure of the bulk material, the shell, which constitutes a thin layer that is chemically distinct from the bulk, and the surface layer, which can bind to small molecules, metal ions, surfactants, or polymers. The presence of a free boundary gives the particle inherently different properties. In the case of phonon propagation, it adds additional modes due to surface wave propagation \cite{Sta2024}. If the surface is coated, it can significantly change the phonon propagation. 

Magnetic nanoparticles, such as magnetite (\ce{Fe3O4}) have additional properties governed by their size, making them practical in many applications such as spintronics \cite{Ansari2021, Li2013, Liu2015}, magnetic hyperthermia \cite{Das2016, Yang2015, Kolenko2014}, drug delivery \cite{Nordin2023, Dorniani2012, Yew2020} and environmental remediation \cite{Koli2019, Goyal2018, Liu2023}. For this reason, it is critical to understand the behavior of phonons under different conditions.

In this article, we characterize the phonon density of states of \ce{Fe3O4} nanoparticles and thoroughly examine several factors that influence the phonon density of states, including nanoparticle size, temperature, particle clusters, and the addition of surface water. The magnetic properties are not considered in this work. 

The manuscript is organized as follows. Section \ref{sec: methods} describes the numerical setup and techniques. The results are presented in Section \ref{sec: results}, where we show the force field, nanoparticle size, and temperature dependencies. We also explore the effects of a cluster of particles and the inclusion of water on the surface of the particles. The appendices present the convergence test, force field parameters, and bulk water simulation. Conclusions and discussions are given in Section \ref{sec: conclusions}.

\subsection{Notation and abbreviations}
\label{subsec: notation}

We used two molecular dynamics (MD) software. The \textsc{Large-scale Atomic/Molecular Massively Parallel Simulator} (LAMMPS)\cite{LAMMPS2022} and GROMACS \cite{GROMACS2020}. We calculated the phonon density of states (PDOS) or vibration density of states (VDOS), eccentricity, and root mean square deviation (RMSD) with the \textsc{Molecular Dynamics Analysis for Neutron Scattering Experiments} (MDANSE) \cite{MDANSEref}. We performed density functional theory (DFT) calculations using the \textsc{Quantum Espresso} (QE)\cite{GiaBarBon2009, GiaAndBruBunBuo2017, GiaBasBon2020}. The DFT phonon calculation was performed using the finite displacement method (FDM) with the help of \textsc{Phonopy} \cite{Togo2023ImplementationPhono3py, Togo2023First-principlesPhono3py}. The PDOS analysis was done with \textsc{Euphonic} \cite{FaiJacVon2022}. We used isothermal–isobaric (NPT) and canonical (NVT) ensembles. We compared our results with time-of-flight inelastic neutron scattering (TOF-INS). 

\section{Methods}
\label{sec: methods}

We used a custom Python script \cite{GalavizGit2025} to build the \ce{Fe3O4} nanoparticles. The initial structure was retrieved from the Crystallography Open Database \cite{GraChaDow09}. We use a \ce{Fe3O4} structure with ID 9006189 \cite{NeiDol94}. Magnetite is a metal oxide with an inverse spinel structure with \ce{Fe3+} in the tetrahedral sites and a mixture of \ce{Fe2+} and \ce{Fe3+} at a ratio of 1:1 in the octahedral sites. The script makes a supercell large enough to contain the desired nanoparticle. Then, it removes the atoms outside a sphere of a given radius while ensuring a neutral structure is achieved. We note that physical magnetite nanoparticles are primarily composed of a \ce{Fe3O4} core with a shell made primarily of $\gamma$-\ce{Fe2O3}. For this work, we will simulate spheres of \ce{Fe3O4}. A more realistic but complex nanoparticle structure is left for future work. The script outputs the structure in LAMMPS and GROMACS format. 

We simulated particles of radius 1, 2, 4, and 8 $nm$ and tested the effect of the numerical domain size. The particle rotates when embedding the nanoparticle in a large numerical domain. The rotation is likely due to the thermostat since it changes randomly over time, and we enforce zero initial translational and angular momentum. This rotational motion produces an elastic-like peak in the PDOS and an oscillatory curve in the RMSD. The effect disappears when the size of the numerical domain fits the nanoparticle. Since the PDOS does not significantly change except for the peak at \qty{0}{meV}, all the simulations were done using the smallest possible numerical domain. 

We tested several force fields. For LAMMPS, we implemented three \texttt{reaxff} potentials \cite{ff_RFF1, ReaxFF2022}, one Buckingham potential \cite{ff_CB1}, and one Lennard-Jones \cite{ff_LJ}. For GROMACS, we used the potential from the reference \cite{ff_LJ} and a TIP4P2005 flexible water model \cite{TIP4P2005f}.  

We performed production runs at several temperatures, progressively heating the system using the following simulation protocol:  
\begin{enumerate} 
    \item  Initial atom relaxation with tolerance \qty{1000}{kJ.mol^{-1}.nm^{-1}}.
    \item  Configuration starts at $T_i$, generating a Gaussian velocity distribution that neutralises total angular and linear momentum.
    \item  $NVT$ heating from $T_n$ to $T_{n+1}=T_n+\Delta T$ for \qty{20}{ps}. \label{itm: restart}
    \item  $NVT$ equilibration for \qty{20}{ps}.
    \item  $NVT$ production for \qty{2500}{ps}. 
    \item  $NVT$ phonon sampling for \qty{50}{ps}. 
    \item  Set $n \rightarrow n+1$. If $T_n<T_f$, continue to \ref{itm: restart}.
    \item  $NPT$ cooling from $T_f$ to $T_i$ for \qty{10}{ps} at \qty{0}{bar}. 
\end{enumerate}

The simulation time, timestep, and sampling output were determined by performing convergence tests (see \ref{app: convergence_test}). Simulations were performed at \qty{200}{K} and \qty{300}{K}. We present the majority of results at \qty{200}{K} since the PDOS is not dramatically sensitive to temperature changes (see Section \ref{subsec: temperature_dependency}). 

The production run has two stages. The first generates a sampling for testing stability, and the second is for phonon calculations. For this system, the phonon energy range drops to zero for energies larger than $E=\qty{150}{meV}$. We calculated phonons up to $E_{max}=\qty{200}{meV}$. Using the equivalence $ E=\hbar \omega$ with $\hbar \approx\qty{0.6582}{meV.ps}$, it is possible to calculate the required output frequency $\omega=\qty{455.78}{rad.ps^{-1}}$. MDANSE calculates a symmetric energy range. Therefore, we need to divide by a factor of two in the conversion. The sampling interval required for the calculation is $\delta t=\pi\omega^{-1}\approx \qty{0.01}{ps}$. From our test, we found that 5000 sampling frames are enough for PDOS calculations. Therefore, a \qty{50}{ps} simulation was used for phonon calculations and \qty{2500}{ps} for stability. We required \qty{10}{ps} for heating and equilibration.

\subsection{Force fields}
\label{subsec: force_fields}
We tested three types of force fields, which we will refer to according to the ID denoted in Table \ref{tab: forcefields}.
\begin{table}[h]
    \centering
    \begin{tabular}{cccc}
         ID & Type & Reference &Comment\\
         \hline\hline
         LJ  & Lennard-Jones & \cite{ff_LJ} &\\
         CB & Buckingham & \cite{ff_CB1} &\\
         ReaxFF2010-full & ReaxFF & \cite{ff_RFF1} & Full\\
         ReaxFF2010-ox & ReaxFF & \cite{ff_RFF1} & Oxides\\
         ReaxFF2022 & ReaxFF & \cite{ReaxFF2022} &\\
    \end{tabular}
    \caption{Type of each potential and reference. We will use ID to refer to each potential. The \textit{Oxides} was trained with a subset of the \textit{Full} training data (see main text for details).}
    \label{tab: forcefields}
\end{table}

The Lennard-Jones (LJ) type combines a potential given by:
\begin{equation}
    V_{\mathrm{LJ}}(\vec{r}_{ij}):=4 \epsilon_{ij} \left [ \left( \frac{\sigma_{ij}}{r_{ij}}\right)^{12} - \left( \frac{\sigma_{ij}}{r_{ij}}\right)^{6} \right], \label{eq: LJ}
\end{equation}
and a Coulomb potential:
\begin{equation}
V_{\mathrm{c}}(\vec{r}_{ij}):=\frac{q_iq_j}{\epsilon r_{ij}}, 
\end{equation} \label{eq: Coulomb}
where $\vec{r}_{ij}$ is the separation between atom $i$ and $j$, $r_{ij}=\vert \vec{r}_{ij}\vert $ is the distance between the atoms, $\epsilon$ is the dielectric constant, $\sigma_{ij}$ and $\epsilon_{ij}$ are parameters, $q_i$ and $q_j$ are the electric charges. The LJ parameters were adapted from the ClayFF \cite{ClayFF2004, ClayFF2021} (see Table \ref{tab: LJ_parameters}). The ClayFF parameters are calibrated to reproduce observed structural and physical properties of materials. 

The Buckingham type combines the Coulomb potential \eqref{eq: Coulomb} with the following potential
\begin{equation}
    V_{\mathrm{BC}}(\vec{r}_{ij}):= A \mathrm{e}^{-r_{ij}/B}-\frac{C}{r_{ij}^6}.
\end{equation}
In this case, the force field parameters are $A$, $B$, and $C$. In \cite{ff_CB1}, the authors fitted the parameters based on quantum mechanical calculations, lattice features, and elastic properties observed from experiments. Table \ref{tab: CB_parameters} shows the corresponding values.

The ReaxFF \cite{ReaxFF2001} is a bond-order-based family of potentials, allowing for continuous bond formation and breaking. The potential has the following functional form: 
\begin{equation}
\begin{array}{lcl}
    V_{\mathrm{reaxff}}(\vec{r}_{ij})&=&V_{bond}(\vec{r}_{ij})+ 
    V_{\mathrm{over}}(\vec{r}_{ij})+
    V_{\mathrm{under}}(\vec{r}_{ij})+\\
    &&V_{\mathrm{val}}(\vec{r}_{ij},\vec{r}_{jk},\Theta_{ijk})+
    V_{\mathrm{pen}}(\vec{r}_{ij},\vec{r}_{jk})+\\
    &&V_{\mathrm{tors}}(\vec{r}_{ij},\vec{r}_{jk},\vec{r}_{kl},\Theta_{ijk},\Theta_{jkl})+\\
    &&V_{\mathrm{conj}}(\vec{r}_{ij},\vec{r}_{jk},\vec{r}_{kl},\Theta_{ijk},\Theta_{jkl})+\\
    &&V_{\mathrm{vdWaals}}(\vec{r}_{ij})+
    V_{\mathrm{Coulomb}}(\vec{r}_{ij}),    \label{eq: reaxff}
\end{array}
\end{equation}
where the right-hand side terms are the bond energy, over-coordination correction, under-coordination correction, valence angle terms, the torsion angles energy, conjugation effects, van der Waals and Coulomb interactions, respectively. The indices $i,j,k,l$ denote four distinctive atoms. The angle between atoms $i,j$ and $k$ is denoted by $\Theta_{ijk}$. The functional forms and parameters are given in \cite{ReaxFF2001}. Regarding functional form and computational overhead, the ReaxFF is significantly more complex than Lennard-Jones or Buckingham. It requires hundreds of parameters. The parameters are fitted using DFT calculations and experimental data. Typically, a ReaxFF provides quantum mechanical accuracy and classical MD scalability.  

For ReaxFF2010, the \textit{Oxides} variant was trained using data that included: two deprotonation reactions of \ce{Fe2+} and \ce{Fe3+} in
solution, dissociation of \ce{Fe}-dimer in solution, heats of formation of hematite, goethite, lepidocrocite, akagan\'eite, w\"ustite, magnetite, and elastic responses of hematite and goethite. The force field parameters of the \textit{Full} case are based on the \textit{Oxides} sets and previous training for calculations on single-\ce{Fe} containing compounds with oxygen, \ce{OH}, and \ce{H2O}.

The ReaxFF2022 was parametrised using DFT calculations and experimental data. The authors improved the ReaxFF2010 by optimising the \ce{Fe}-\ce{Fe}, \ce{Fe}-\ce{O}, and \ce{Fe}-\ce{H} parameter sets.

\subsection{Analysis}
\label{subsec: analysis}

 Regarding the PDOS, each sampling of LAMMPS and GROMACS trajectories consisted of 5000 position snapshots spaced with steps of \qty{0.01}{ps}. For LAMMPS, we use second-order interpolation to generate the velocities. GROMACS does not require velocity calculation since the output file includes the velocity. The total PDOS was weighted using the \texttt{b\_incoherent} and \texttt{b\_coherent} options. The PDOS is normalized by the total degrees of freedom:
\begin{align}
    \mathrm{PDOS}(E)=&3N\frac{\mathrm{pdos}(E)}{||\mathrm{pdos}(E_{\mathrm{max}})||}, \label{eq: dos_normalization}\\
        ||\mathrm{pdos}(E_{\mathrm{max}})||:=&\int_{0}^{E_{\mathrm{max}}}|\mathrm{pdos}(E)| \mathrm{d}E, \nonumber
\end{align}
    
where $\mathrm{pdos}(E)$ is the PDOS as calculated by MDANSE and $N=137$ is the total number of atoms in the  \ce{Fe3O4} unit cell. We used \eqref{eq: dos_normalization} when comparing PDOS from MD simulations, as the PDOS is calculated by the same software and under equivalent conditions. 

We also normalize the PDOS by its maximum intensity:
\begin{align}
    \mathrm{PDOS}(E)=&\frac{\mathrm{pdos}(E)}{\lceil{\mathrm{pdos}(E_{\mathrm{max}})\rceil}}. \label{eq: dos_max_normalization}\\
    \lceil{\mathrm{pdos}(E_{\mathrm{max}})}\rceil:=&\max_{E\in [0,E_{\mathrm{max}}]}\left [\mathrm{pdos}(E)\right]. \nonumber
\end{align}
 We used \eqref{eq: dos_max_normalization} when comparing with experimental and DFT calculations, as the PDOS is calculated using different methods, and the resulting PDOS is more convenient for visualization.  

We calculated the PDOS relative difference as a function of size and temperature given by:
\begin{equation}
\mathrm{PDOS}_{\mathrm{diff}}(x_1,x_2):=100\frac{|| \mathrm{PDOS}(E_{\mathrm{max}};x_1)-\mathrm{PDOS}(E_{\mathrm{max}};x_2) ||}{\frac{1}{2}|| \mathrm{PDOS}(E_{\mathrm{max}};x_1)+\mathrm{PDOS}(E_{\mathrm{max}};x_2)||}, \label{eq: relative_difference}
\end{equation}
where $x_1,x_2$ correspond to two different temperatures or particle sizes. 

 The radius of gyration was calculated using a custom C++ program that reads both LAMMPS and GROMACS trajectories and computes the following:
\begin{equation}
    R_g = \left ( \frac{\sum_i m_i \| \vec{r}_i-\vec{r}_{\mathrm{com}} \|^2}{\sum_i m_i} \right)^{\frac{1}{2}},
\end{equation}
where $m_i$ is the mass of atom $i$, $\vec{r}_i$ its position and $\vec{r}_{\mathrm{com}}$ the center of mass of the system.
In the case of GROMACS, the output was validated using GROMACS' \texttt{gyrate} and \texttt{rmsdist} commands.  
The same software calculates the radial distribution histogram. The program calculates the center of mass for each trajectory frame and computes the distance of each atom to the respective center of mass. The histogram aggregates every calculated distance for every atom at every trajectory frame. The result accounts for every radial location visited by each species during the simulation. 

The structures were visualized using the VMD software \cite{HumDalSch1996}.

\subsection{Density functional theory calculation}

Initial atomic positions from the \ce{Fe3O4} structure were used to generate input files for QE. The files were generated using the Materials Cloud
platform \cite{TalKumPasYakGra2020}. We performed an energy cutoff test (\texttt{ecutwfc} parameter in QE) with energies between \qty{60}{Ry} and \qty{110}{Ry}. We selected \ce{90}{Ry} as the optimal value with less than 0.001\% relative change to the calculated energy at 110 Ry. Similarly, we performed a k-point convergence test, choosing a $9\times 9 \times 9$ mesh without offset. We used those parameters for relaxation and phonon calculations. We performed variable cell relaxation using damped (quick-min Verlet) dynamics, a 0.0001 force convergence threshold (\texttt{forc\_conv\_thr}) and a 0.0014 energy convergence threshold (\texttt{etot\_conv\_thr}). For phonon calculation, we generated the displacement structures using \textsc{Phonopy} on a $2\times 2 \times 2$ supercell, resulting in four perturbed structures. The PDOS was calculated from \qty{0}{meV} to \qty{150}{meV} with a \qty{0.5}{meV} resolution and \qty{5}{meV} Gaussian broadening width.

\section{Results}
\label{sec: results}

\subsection{Force field dependency}
\label{subsec: force_field_dependency}

The force field, which describes interactions within and between atoms in molecules, is crucial for properly characterizing \ce{Fe4O3} nanoparticles. We tested several force fields (see Table \ref{tab: forcefields}). Each force field was evaluated by comparing the calculated PDOS derived from the trajectories with nuclear resonant scattering of synchrotron radiation results from \cite{SetKitKob2003}. The experimental phonon data includes all \ce{Fe} sites in \ce{Fe3O4}. The measured sample was polycrystalline magnetite (\ce{Fe3O4}) pressed into a pellet.

\begin{figure}[!ht]
\centering
    \includegraphics[width=0.75\textwidth]{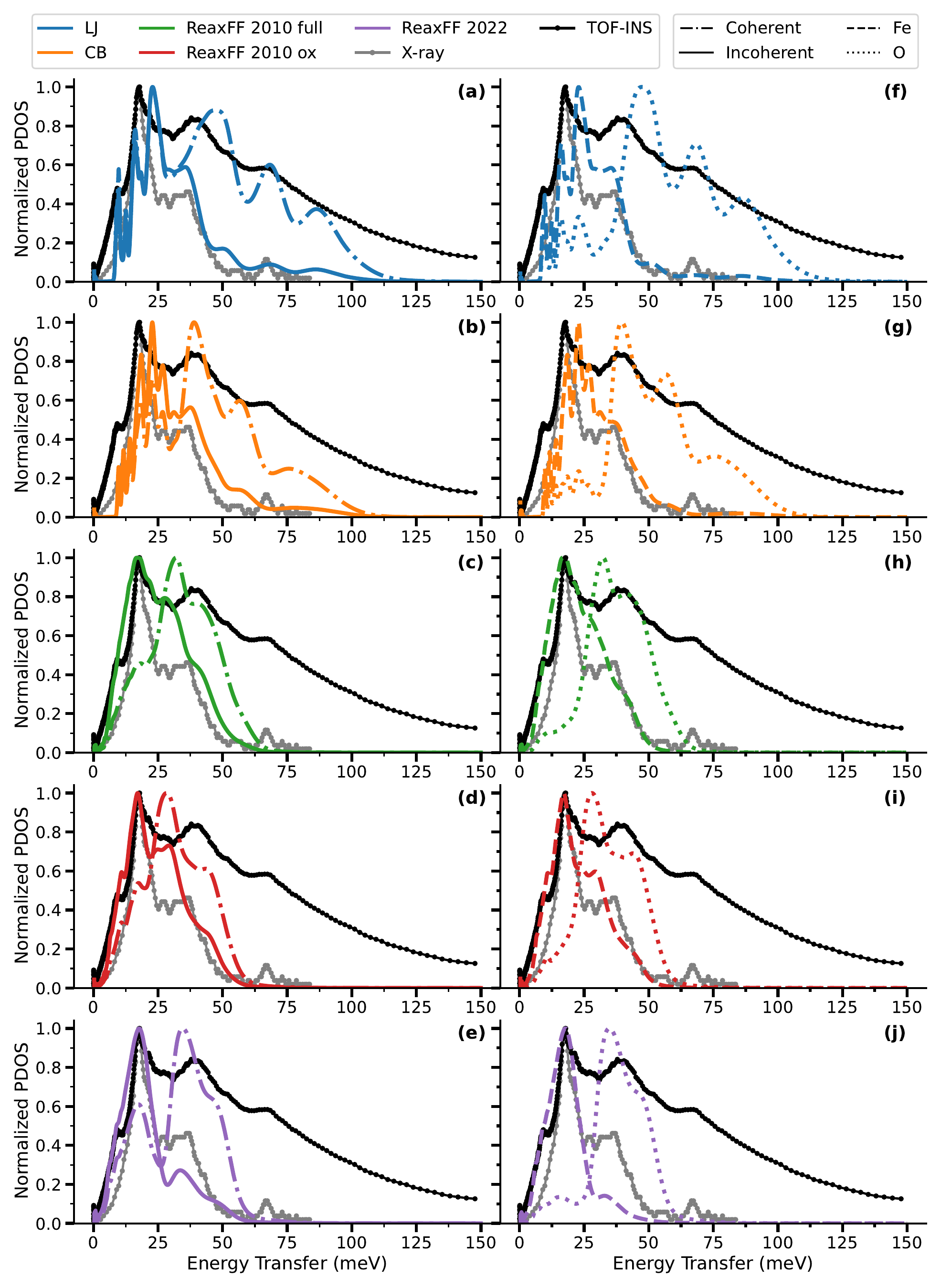}
\caption{PDOS force field comparison. The results are from a $L=\qty{2}{nm}$ supercell at \qty{200}{K}. Panels \textbf{(a)}-\textbf{(e)} show the total density of state weighted by coherent (solid line) and incoherent (dashed-dotted line) scattering for each tested force field. Panels \textbf{(f)}-\textbf{(j)} show the corresponding partial density of state for \ce{Fe} (dashed line) and \ce{O} (dotted line). Each panel's grey and black lines correspond to the experimental X-ray and neutron scattering PDOS, respectively. Each PDOS was normalised by its maximum intensity.} \label{fig: ff_dos}
\end{figure}

In addition, we compare the PDOS with time-of-flight inelastic neutron scattering (TOF-INS) data of a magnetite $r=\qty{100}{nm}$ particle at \qty{500}{K}, to remove the magnon contribution. The TOF-INS data were collected on the Pelican instrument at the Australian Nuclear Science and Technology Organization (ANSTO) \cite{Pelican1, Pelican2}. Pelican was operated with a neutron wavelength of $\lambda=\qty{4.69}{\angstrom}$, which afforded an energy resolution at the elastic line of \qty{0.13}{meV}. Resolution increases gradually with increased energy transfer according to the resolution function \cite{Pelican1, Pelican2}. Around \qty{12}{g} of sample was placed in an annular \ce{Al} can with an overall neutron path length of \qty{4}{mm} to reduce multiple scattering events. Data were collected for 4 hours at \qty{500}{K}. The data were corrected using an empty-can background subtraction, and detector efficiencies were corrected by normalizing to a vanadium sample scan. The Large Array Manipulation Program (LAMP) \cite{LAMP} was used for data reduction, analysis, and derivation of PDOS. A detailed analysis of the experimental data is given in \cite{PorGalYu2025}. 

Figure \ref{fig: ff_dos} presents the PDOS comparison between experimental data and different force fields. Given that the X-ray experimental data is from a bulk sample, we performed the comparison with a periodic supercell of size $L=\qty{2}{nm}$. We will explore the particle size effect later (see Section \ref{subsec: size_dependency}). We performed two sets of simulations, one at \qty{200}{K} and the other at \qty{300}{K}. The results have no significant difference (see Section \ref{subsec: temperature_dependency}). We focused our analysis on the acoustic and optic phonon region of up to \qty{100}{meV} since at this temperature range the TOF-INS data include magnetic contribution, which is not included in our modeling. The data was normalised to the maximum intensity of each curve. The LJ and CB force fields' \ce{Fe} peak is shifted to slightly higher energy, but the width is comparable with the experimental result. In the LJ force field result, a secondary peak at around \qty{67}{meV} coincides with an \ce{O} peak also present in the TOF-INS data. The three ReaxFF's \ce{Fe} results reproduce the main peak at the correct energy; however, the \ce{O} is located at a lower energy.

\begin{figure}[!ht]
\centering
    \includegraphics[width=1\textwidth]{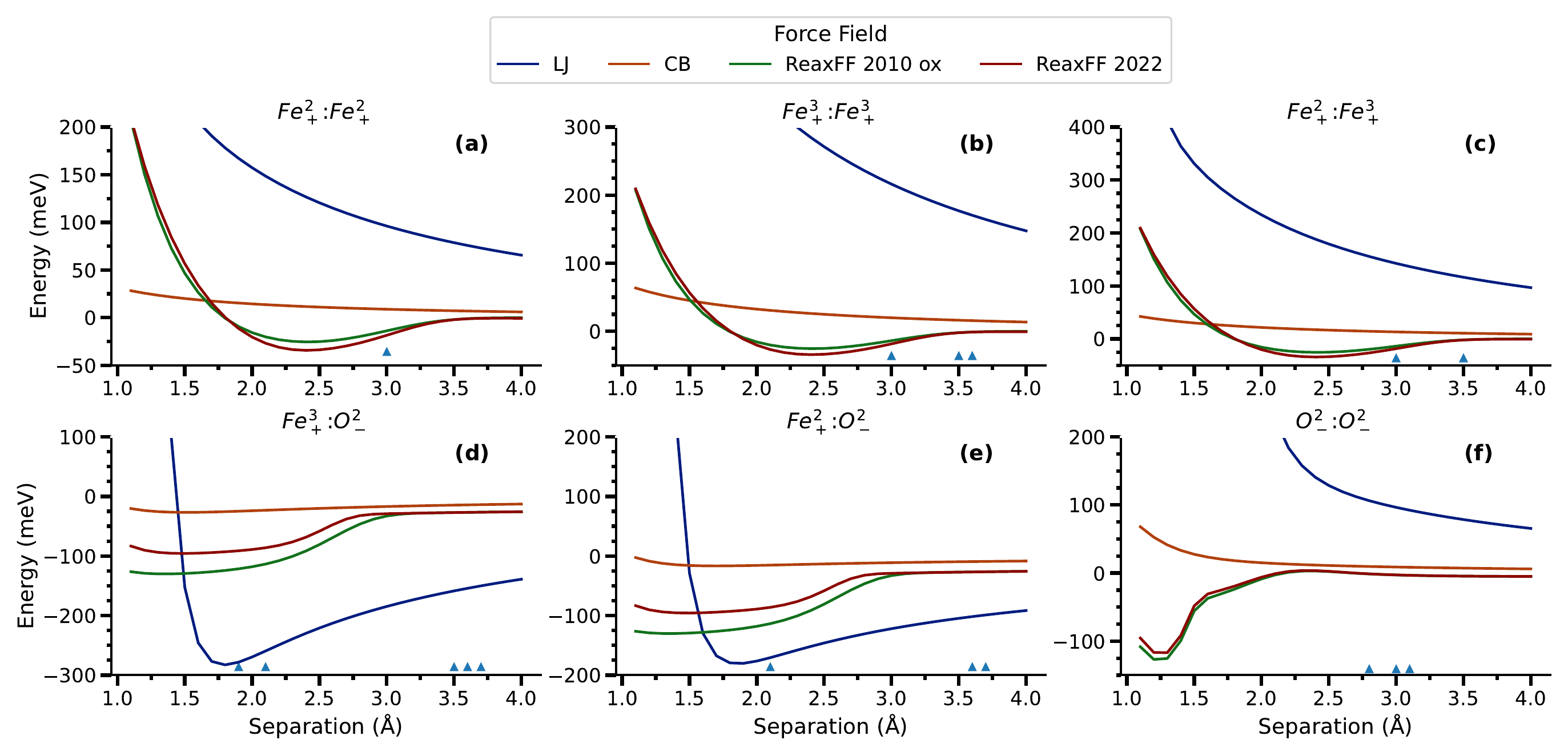}
\caption{Pair energy for each force field. Panels \textbf{(a)} to \textbf{(f)} show the energy calculated by LAMMPS for each species pair in isolation at several separations. The triangle symbol indicates the nearest neighbours as given in the initial \ce{Fe3O4} structure file.} \label{fig: potential}
\end{figure}

Figure \ref{fig: potential} shows a sampling of the pair energy for the six possible species pairs. We generated the curves by calculating the pair energy for two atoms in a large cubic domain for several separation values. The results show that the LJ and CB force fields are repulsive between species of the same type (see Figure \ref{fig: potential} \textbf{(a), (b), (c)} and \textbf{(f)}.), and attractive between \ce{Fe} and \ce{O}. In contrast, the ReaxFF produces an attractive force for every pair except for \ce{O}-\ce{O} pairs. This counterbalanced behavior explains the larger spread of energies in the phonon density of states for LJ and CB. However, we should notice that the ReaxFF behavior is more complex, since it takes additional terms beyond pair interactions (refer to Eq. \eqref{eq: reaxff}). Overall, the five tested force fields produce a relatively good PDOS for energy below \qty{100}{meV}. For higher energy, three factors affect the modeling: the presence of hydrogen, high-energy magnetic excitation, and instrument resolution broadening. Therefore, we will focus our analysis on energy below \qty{100}{meV}. The LJ force field will be used in most subsequent results since it reproduces both experimental results well and is computationally efficient. We will also present additional results using ReaxFF2022.  

\begin{figure}[!ht]
\centering
    \includegraphics[width=0.85\textwidth]{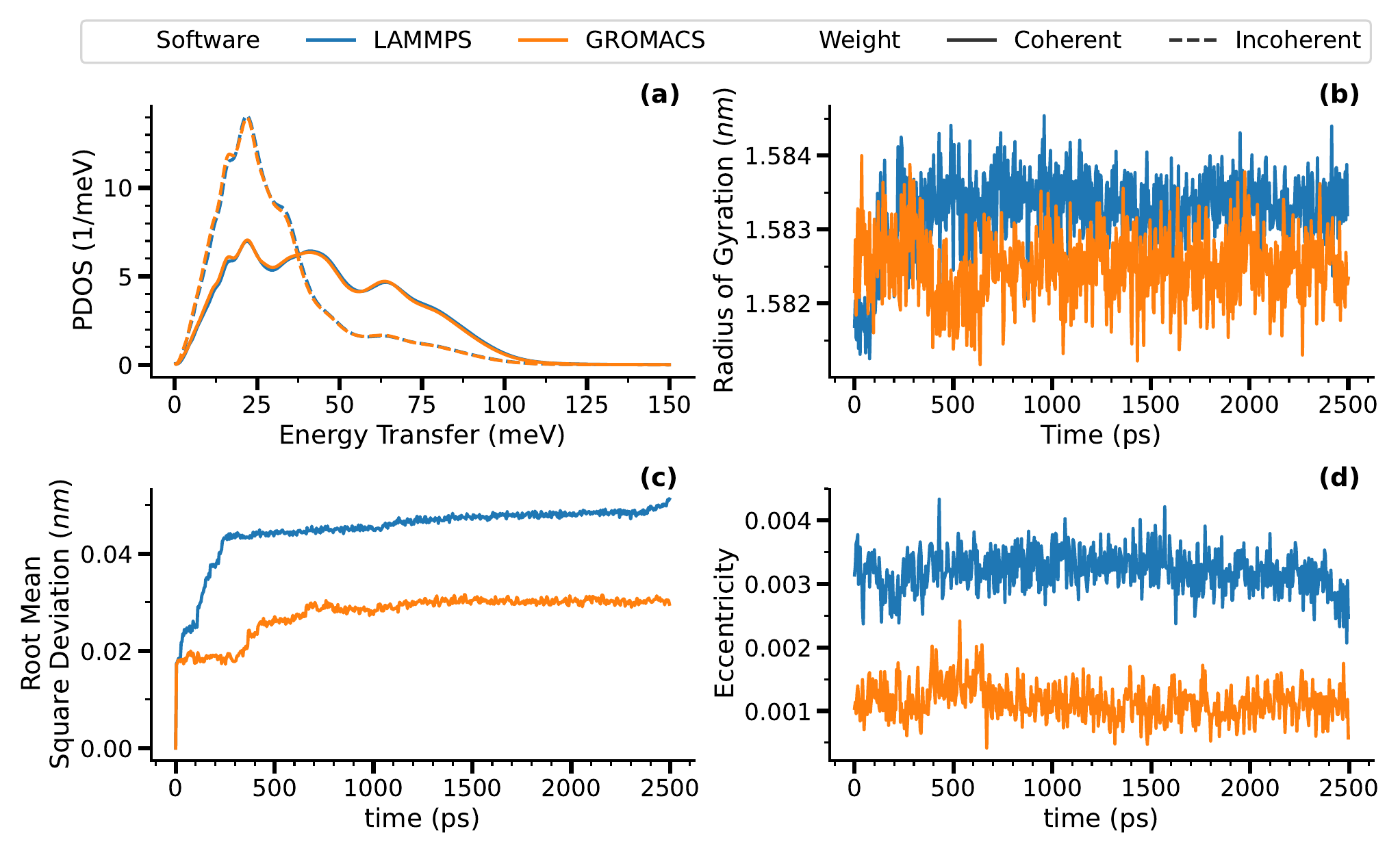}
\caption{Simulation software comparison. The results are from a $r=\qty{2}{nm}$ particle at \qty{200}{K} using the LJ force field. Panel \textbf{(a)} shows the total density of state weighted by coherent and incoherent scattering for each tested software. Panel \textbf{(b)} shows the corresponding radius of gyration. Panel \textbf{(c)} shows the root mean square deviation. Panel \textbf{(d)} shows the eccentricity.} \label{fig: lammps_gromacs_LJ}
\end{figure}

 Figure \ref{fig: lammps_gromacs_LJ} presents the results of implementing the LJ force field in GROMACS. There are only minor differences between the calculated PDOS of LAMMPS and GROMACS (see Figure \ref{fig: lammps_gromacs_LJ}-\textbf{(a)}). The radius of gyration is approximately \qty{1.58}{nm} in both cases (Figure \ref{fig: lammps_gromacs_LJ}-\textbf{(b)}), and the root mean square deviation shows a comparatively small deviation from the initial structure (Figure \ref{fig: lammps_gromacs_LJ}-\textbf{(c)}). The eccentricity (Figure \ref{fig: lammps_gromacs_LJ}-\textbf{(d)}) is approximately zero in both cases. The almost constant radius of gyration and eccentricity imply that the particle does not significantly change in size or shape during the simulation. Overall, we obtained comparable results independent of the software used. In the following sections, we will show results generated with GROMACS.

\begin{figure}[!ht]
\centering
    \includegraphics[width=0.85\textwidth]{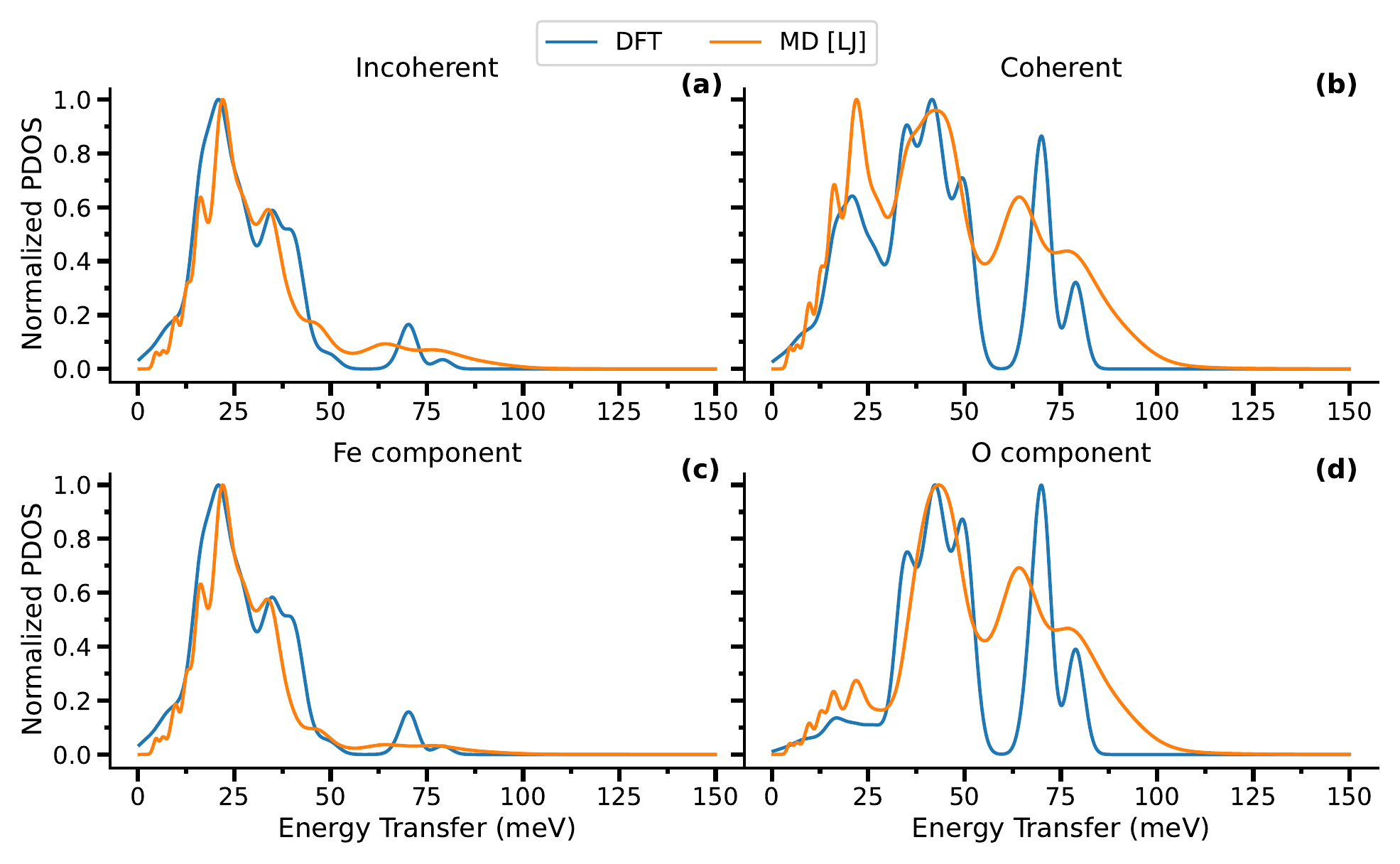}
\caption{Bulk magnetite calculation comparison. The density functional theory calculation is from a finite displacement method using Quantum Espresso. The molecular dynamics result is from a periodic supercell block of size \qty{3.35}{nm} at \qty{200}{K} using the LJ force field. Panel \textbf{(a)} shows the total density of state weighted by incoherent scattering. Panel \textbf{(b)} shows the corresponding coherent PDOS. Panels \textbf{(c)} and \textbf{(d)} show the iron and oxygen components of the PDOS.} \label{fig: md_dft_dos}
\end{figure}

As a final test of the force fields, we compare the results of bulk magnetite with DFT calculations. Figure \ref{fig: md_dft_dos} shows a comparison between the DFT calculated PDOS and MD using the LJ force field. Our DFT result is similar to previous calculations \cite{HanKozPar05} and we observe a good fit between the two methods. The LJ force field gives a comparable \ce{Fe} component without the peak near \qty{67}{meV}. The \ce{O} component shows a similar profile with variation in intensity and peak locations.

\subsection{Size dependency}
\label{subsec: size_dependency}

\begin{figure}[!ht]
\centering
    \includegraphics[width=0.85\textwidth]{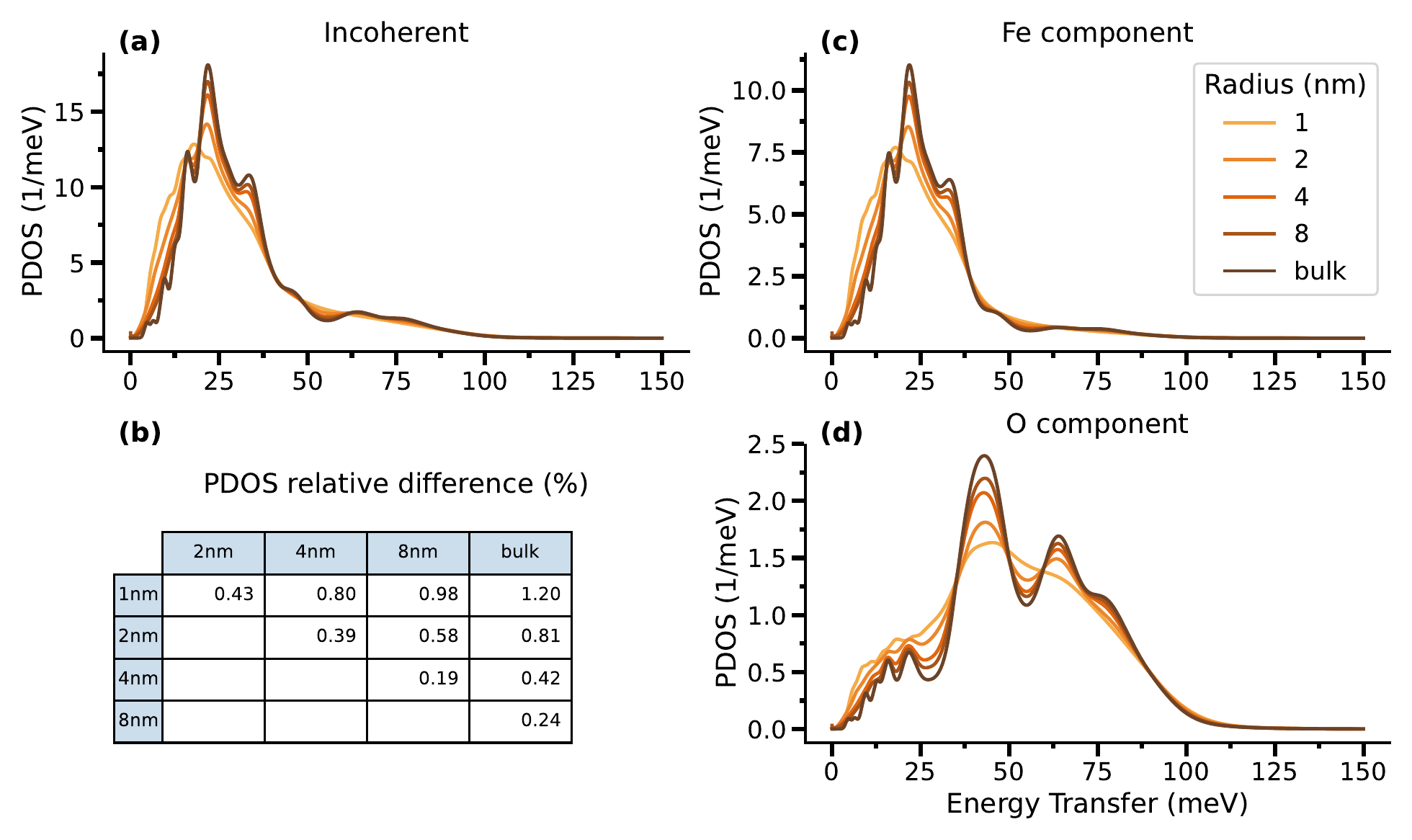}
\caption{PDOS particle size effect. Panel \textbf{(a)} shows the normalized PDOS for different particle sizes. Panel \textbf{(b)} shows the relative difference in percentage given by \eqref{eq: relative_difference}. Panels \textbf{(c)} and \textbf{(d)} show the partial PDOS component for \ce{Fe} and \ce{O}, respectively.} \label{fig: size_comparison}
\end{figure}

Figure \ref{fig: size_comparison} presents a comparison between the PDOS for different particle sizes and a bulk magnetite crystal. As particle size is decreased, the optical phonons significantly broaden due to the increased scattering rate (decreased phonon lifetime) from finite-sized boundaries. This effect scales with the particle size, i.e., the smaller the nanoparticle, the broader the peaks. Phonon softening is observed with decreasing particle size, a phenomenon attributed to enhanced surface strain. Atoms located at the nanoparticle surface, lacking a full complement of nearest neighbors, experience increased strain, which in turn leads to the softening of vibrational modes. As the particle size diminishes, the surface-to-volume ratio increases, resulting in a greater proportion of strained surface atoms and, consequently, a more pronounced softening of the phonon modes. The softening and broadening both seem to become more apparent for higher energy phonon modes. On the low energy side of the PDOS spectrum (between 0 and ~20 meV), there is an increased broad intensity. This is characteristic of Rayleigh surface waves, which have previously been reported in the literature \cite{Sta2024}.

\begin{figure}[!ht]
\centering
    \includegraphics[width=0.85\textwidth]{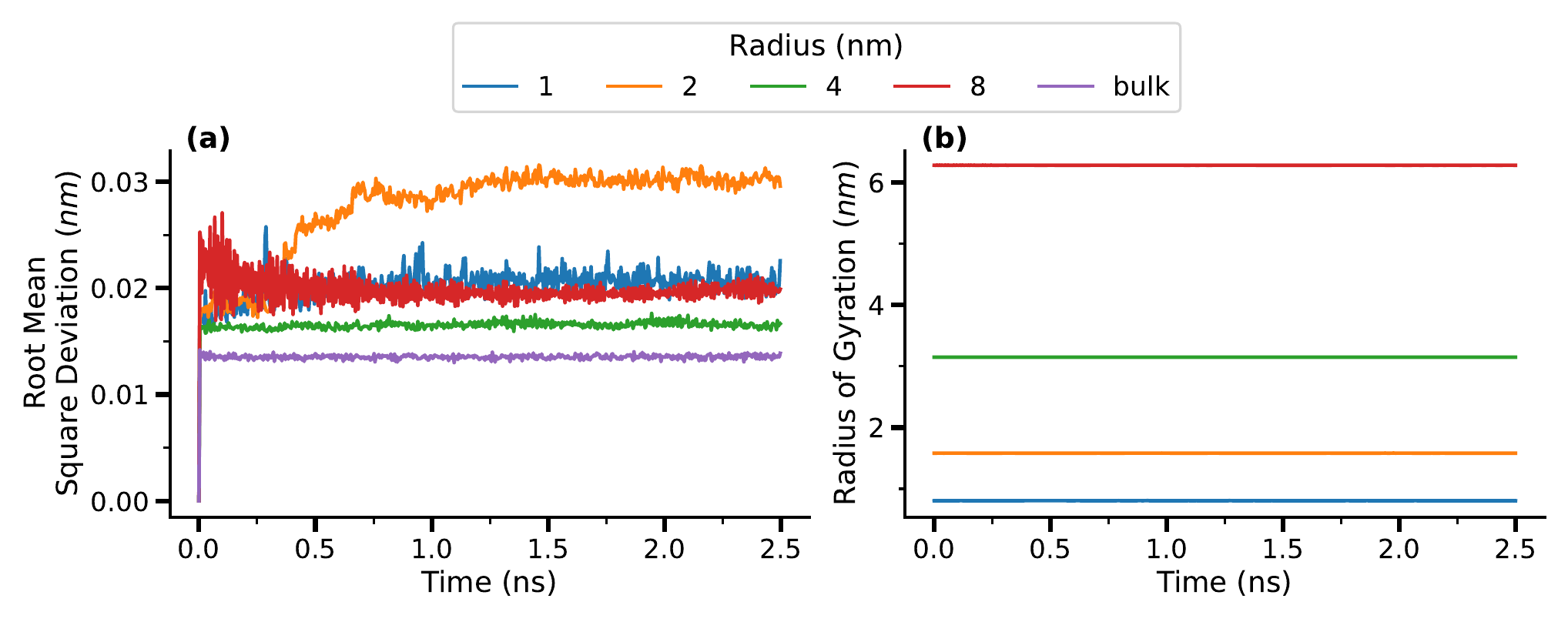}
\caption{Dynamical variables particle size effect. Panel \textbf{(a)} shows the root mean square deviation for different particle sizes. Panel \textbf{(b)} shows the corresponding radius of gyration.} \label{fig: size_dynamics}
\end{figure}

Figure \ref{fig: size_dynamics} shows the root mean square deviation and the radius of gyration. Table \ref{tab: size_dynamics} shows the mean value and standard deviation of the radius of gyration and eccentricity. The results show that there is no significant change in the shape and size of the particle. There is no translation or rotation of the nanoparticle since the numerical domain confines the particle. As mentioned before, we observed nanoparticle rotation due to thermal effects when simulating the particles in a large numerical domain. In that case, the phonon density of states shows a peak at \qty{0}{meV} and a significant variation of the root mean square deviation.

\begin{table}[!ht]
    \centering
    \begin{tabular}{ccc|cc}
        & \multicolumn{2}{c}{Radius of gyration} & \multicolumn{2}{c}{Eccentricity} \\
         Particle size (\unit{nm})& $\mu_{rog}(\unit{nm})$ & $\sigma_{rog}(\unit{nm})$ &$\mu_{ecc}$ & $\sigma_{ecc}$\\
         \hline \hline
            1 & 0.8062 & 0.0007 & 0.0266 & 0.0010 \\
            2 & 1.5825 & 0.0005 & 0.0011 & 0.0003 \\
            4 & 3.1487 & 0.0003 & 0.0118 & 0.0001 \\
            8 & 6.2824 & 0.0006 & 0.0012 & 0.0001 \\                 
    \end{tabular}
    \caption{Radius of gyration and eccentricity mean and standard deviation.}
    \label{tab: size_dynamics}
\end{table}

\subsection{Temperature dependency}
\label{subsec: temperature_dependency}

\begin{figure}[!ht]
\centering
    \includegraphics[width=0.85\textwidth]{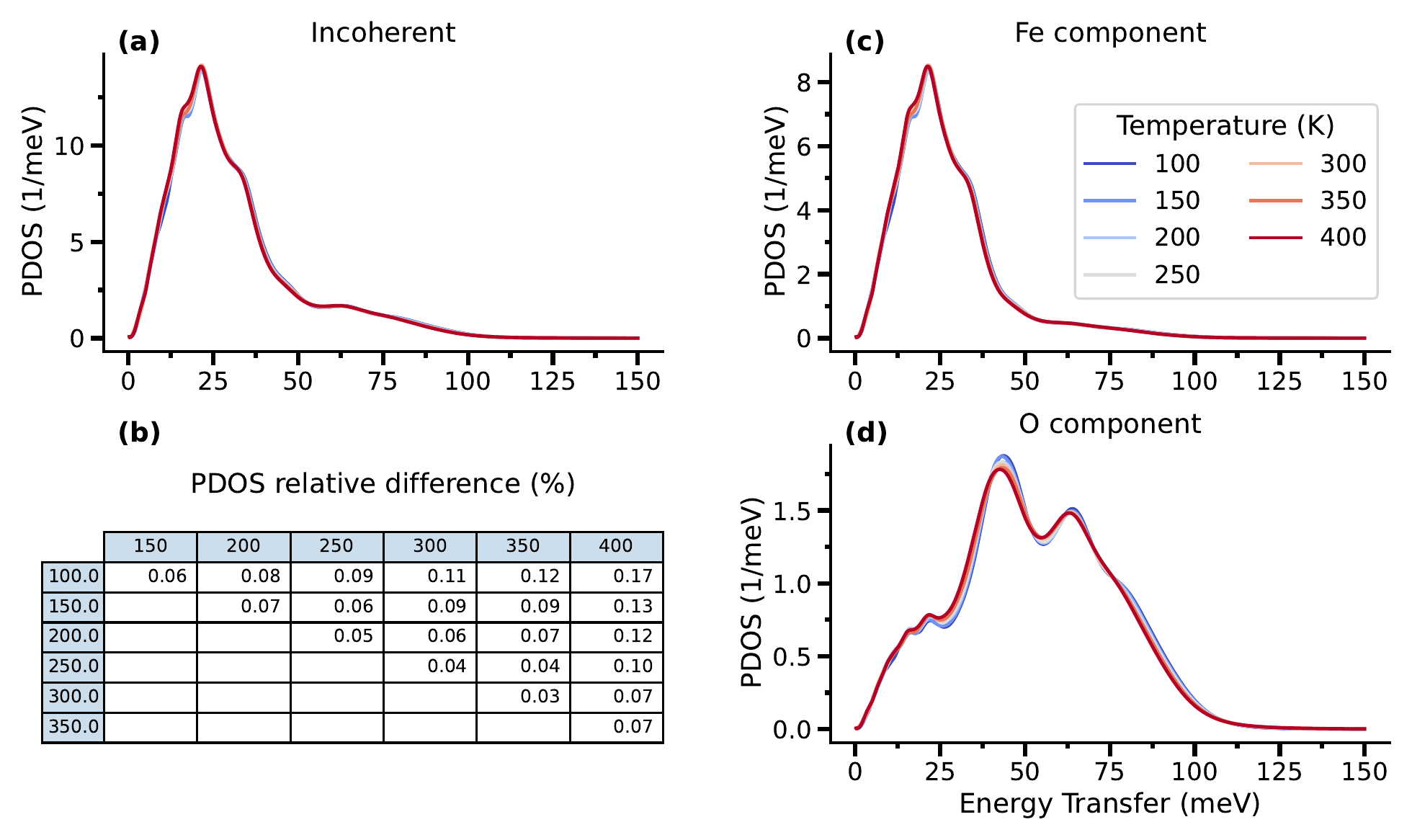}
\caption{Phonon density of states temperature effect. The results are from a $r=\qty{2}{nm}$ particle. Panel \textbf{(a)} shows the PDOS at different temperatures. Panel \textbf{(b)} shows the relative difference in percentage given by \eqref{eq: relative_difference}. Panels \textbf{(c)} and \textbf{(d)} show the partial PDOS for \ce{Fe} and \ce{O}, respectively.} \label{fig: temperature_comparison}
\end{figure}

To explore the temperature effects on the phonon density of states, we simulated a \qty{2}{nm} particle that is progressively heated from \qty{100}{K} to \qty{400}{K} in \qty{50}{K} increments. Figure \ref{fig: temperature_comparison} shows the result. The Fe component shows almost no change upon heating. While the most significant differences are observed in the \ce{O} component, this is only a very slight anharmonic softening. The relative change in the PDOS is less than 1\%. Therefore, we consider the PDOS to be relatively temperature-independent in the simulated temperature range. 

\subsection{Clusters}
\label{subsec: clusters}

\begin{figure}[!ht]
\centering
    \includegraphics[width=0.85\textwidth]{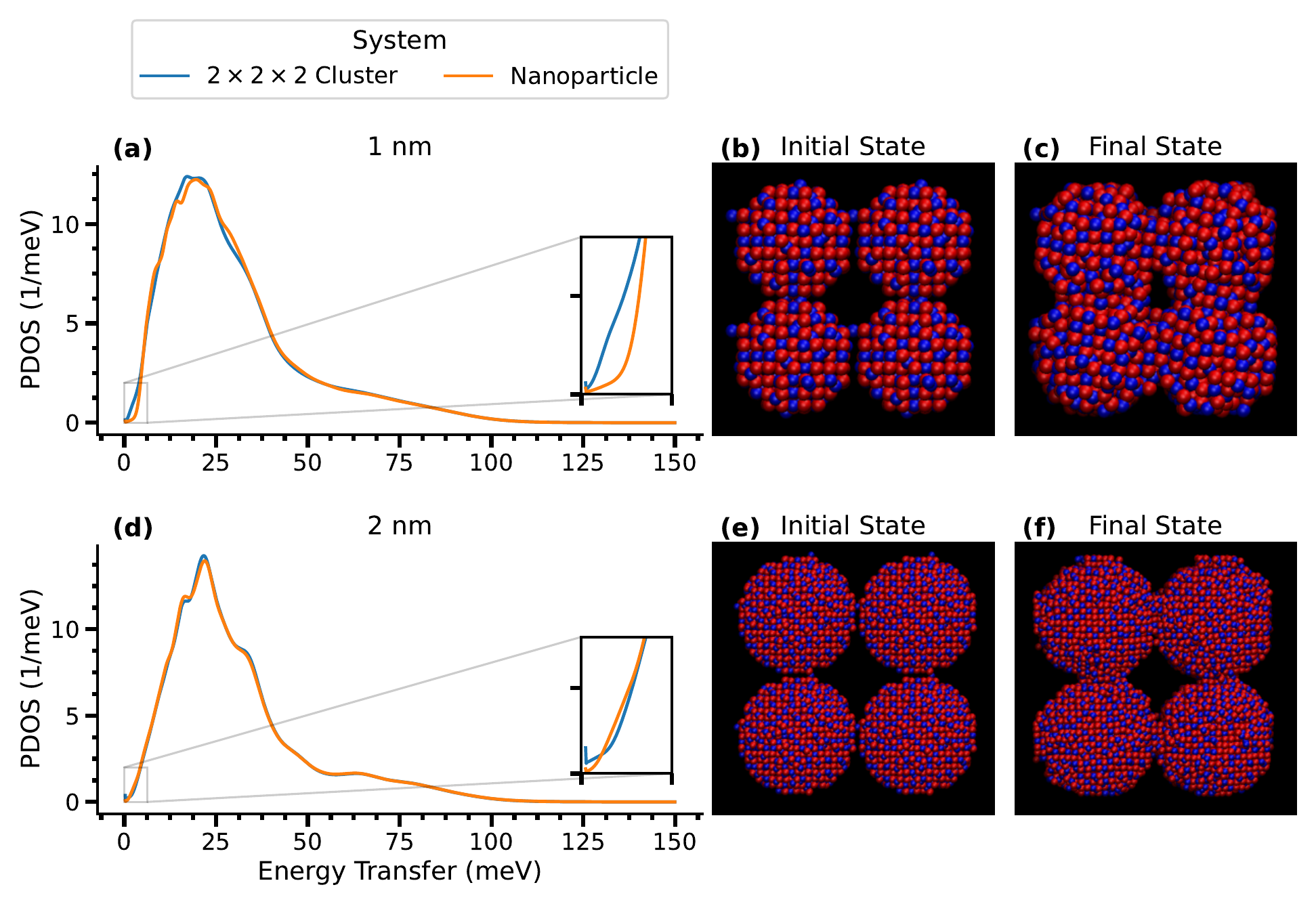}
\caption{Phonon density of states cluster vs isolated particle. The results are from a system at a \qty{200}{K} temperature. Panel \textbf{(a)} shows the incoherent PDOS for a \qty{1}{nm} particle and a $2 \times 2 \times 2$ cluster of the same size particles. Panels \textbf{(b)} and \textbf{(c)} show a side view of the initial and final state of the \qty{1}{nm} particle cluster. Panels \textbf{(d)} to \textbf{(f)} show the corresponding result for \qty{2}{nm} particles.} \label{fig: cluster_particle}
\end{figure}

To explore cluster effects, we simulated multiple nanoparticles. We created two sets of $2 \times 2 \times 2$ clusters (i.e., eight particles in a cubic numerical domain). One using $r=\qty{1}{nm}$ particles and the other with $r=\qty{2}{nm}$ particles. Figure \ref{fig: cluster_particle} shows the result of comparing clusters of particles vs a single particle. In both \qty{1}{nm} and \qty{2}{nm} cases, the resulting coherent PDOS is similar between the single particle and the cluster. However, the low-energy range shows that the \qty{1}{nm} $2\times 2\times 2$ cluster has additional modes not present in the isolated nanoparticle (see inset in Figure \ref{fig: cluster_particle}-\textbf{(a)}). These modes are likely acoustic modes. The isolated particle exhibits a clear acoustic gap, but once multiple particles are combined, the gap begins to close, and collective excitations become possible with longer wavelengths and lower energies. We also looked at the PDOS for \ce{Fe} and \ce{O}, and the differences are minimal as well. This result confirms that the single particle model represents a system with more particles. Panels \textbf{(b)}, \textbf{(c)}, \textbf{(e)} and \textbf{(f)} of Figure \ref{fig: cluster_particle} show a visualization of the initial and final structures at \qty{200}{K}. We observed a slight agglomeration of the particles, which did not significantly alter their structure. Ultimately, the particles maintained their spherical shape after a total simulation time of \qty{5.14}{ns}, which included the transition from \qty{200}{K} to \qty{300}{K}.

\subsection{Surface water}
\label{subsec: surface_water}

\begin{figure}[!ht]
\centering
    \includegraphics[width=0.85\textwidth]{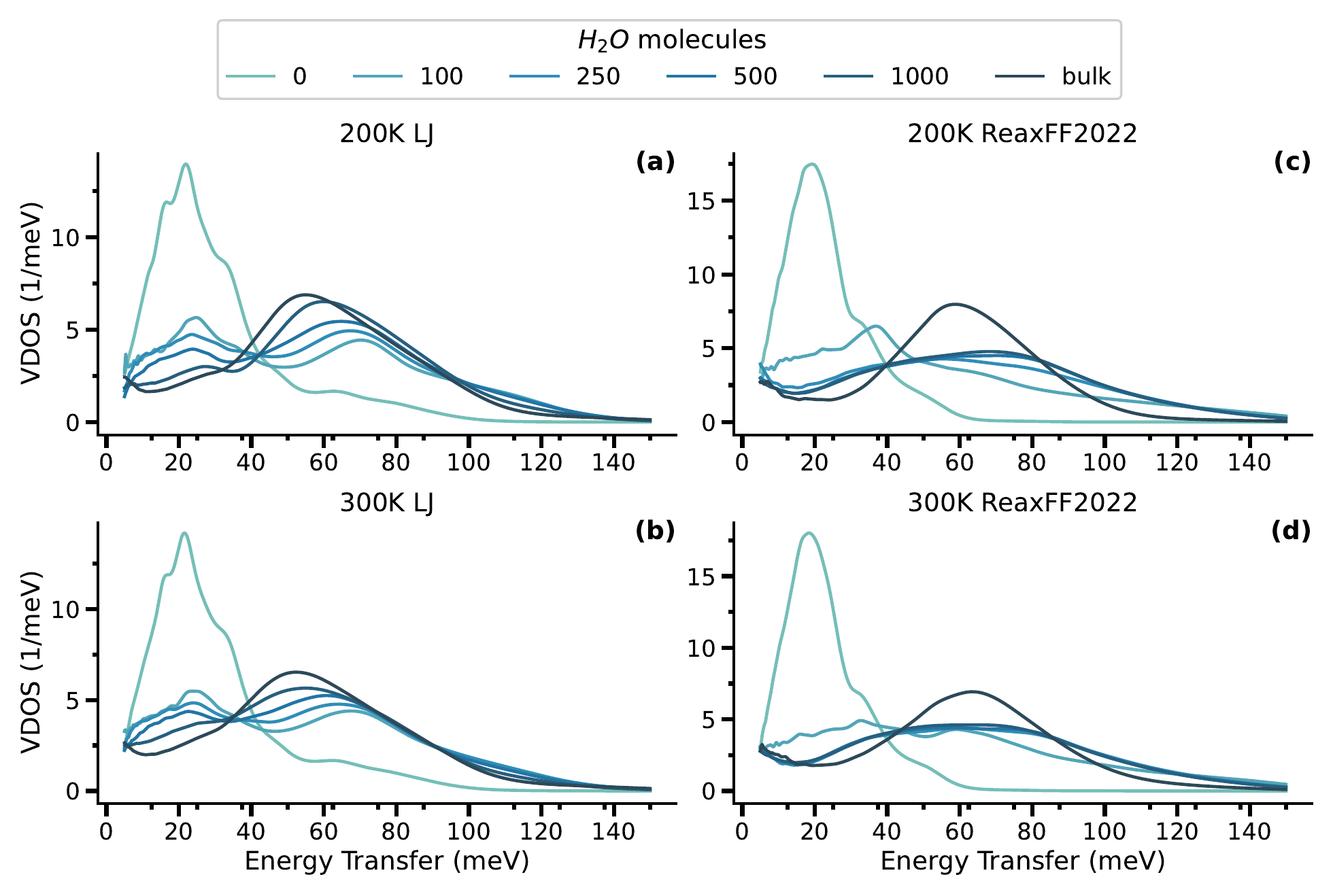}
\caption{VDOS of nanoparticle with water. The results are from a $r=\qty{2}{nm}$ nanoparticle. Panel \textbf{(a)} shows the incoherent VDOS at a \qty{200}{K} temperature using the LJ force field and the TIP4P2005f water model for a particle with 0, 100, 250, 500, and 1000 \ce{H2O} surface molecules. Panel \textbf{(b)} shows the corresponding result for a particle at \qty{300}{K}. Panels \textbf{(c)} and \textbf{(d)} show the results using the ReaxFF2022 force field at \qty{200}{K} and \qty{300}{K}, respectively.} \label{fig: H2O_dos}
\end{figure}

In this section, we look at the influence of surface water on the particle. We simulated a $r=\qty{2}{nm}$ particle and added 100, 250, 500, and 1000 \ce{H2O} molecules. In addition, we simulated bulk water. We used the LJ force field with a TIP4P2005f flexible water model and the ReaxFF2022. Figure \ref{fig: H2O_dos} shows the VDOS\footnote{In this section, we use the term vibrational density of states to account for the added \ce{H2O} molecules located randomly on the nanoparticle's surface.} for a system at \qty{200}{K} and \qty{300}{K} using both force fields. The presence of water significantly changes the vibration density of states even when adding a few molecules. The incoherent VDOS is dominated by the \ce{H} component and shows a prominent peak at energies larger than \qty{50}{meV}. The \ce{O} component includes additional peaks, in this case at low energies. Naturally, the \ce{Fe} is almost unaffected (Figure \ref{fig: H2O_comp_dos} in \ref{app: additional_water_simulation_results} shows the respective components). From our result of bulk \ce{H2O}, we know that the TIP4P2005f water, the peak at around \qty{75}{meV}, is shifted to approximately \qty{50}{meV} (see Appendix \ref{app: bulk_water_simulation}). The simulation using the ReaxFF2022 force field shows a prominent elastic peak that indicates water diffusion even at \qty{200}{K}. In contrast, the LJ simulation only shows diffusion at \qty{300}{K} and for more than 250 \ce{H2O} molecules. The LJ VDOS gradually broadens, transitioning from the VDOS without \ce{H2O} to the bulk VDOS. In contrast, the ReaxFF2024 result appears to indicate three defined states: no water, particle with water, and bulk.   

\begin{figure}[!ht]
\centering
    \includegraphics[width=0.7\textwidth]{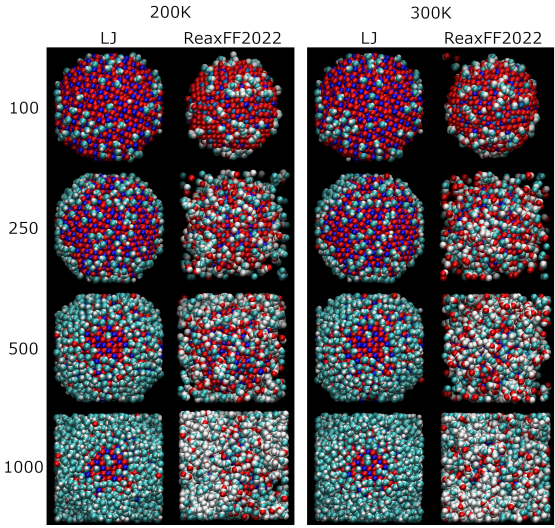}
\caption{Final state of nanoparticle structures with different amounts of water and temperature. The results are from a $r=\qty{2}{nm}$ nanoparticle. Blue spheres represent \ce{Fe} atoms, red spheres are oxygen from \ce{Fe3O4}, cyan spheres are oxygen from \ce{H2O} molecules, and white spheres are hydrogen atoms. Each row shows the nanoparticle with 100, 250, 500, and 1000 \ce{H2O} surface molecules. The first two columns are of a particle at \qty{200}{K} using the LJ and ReaxFF2022, respectively. The last two columns are the corresponding result for a system at \qty{300}{K}.} \label{fig: 2nm_nanoparticle}
\end{figure}

Figure \ref{fig: 2nm_nanoparticle} shows a snapshot of the structures at the final state of the phonon calculation for \qty{200}{K} and \qty{300}{K}. Here we can confirm \ce{H2O} molecules diffusion for almost every simulation using the ReaxFF2022. We observed a reaction between \ce{Fe3O4} and \ce{H2O} with oxygen interchange and the formation of \ce{OH} ions, \ce{O2}, and \ce{H2} molecules. In contrast, simulations performed with the LJ and the TIP4P2005f water model keep their structure with the \ce{H2O} molecules attached to the surface. Notice that the \ce{H2O} molecules were initially randomly distributed in the numerical domain in every simulation. Therefore, the distribution of the \ce{H2O} molecules on the nanoparticle surface is a consequence of the simulation.

\begin{figure}[!ht]
\centering
    \includegraphics[width=0.85\textwidth]{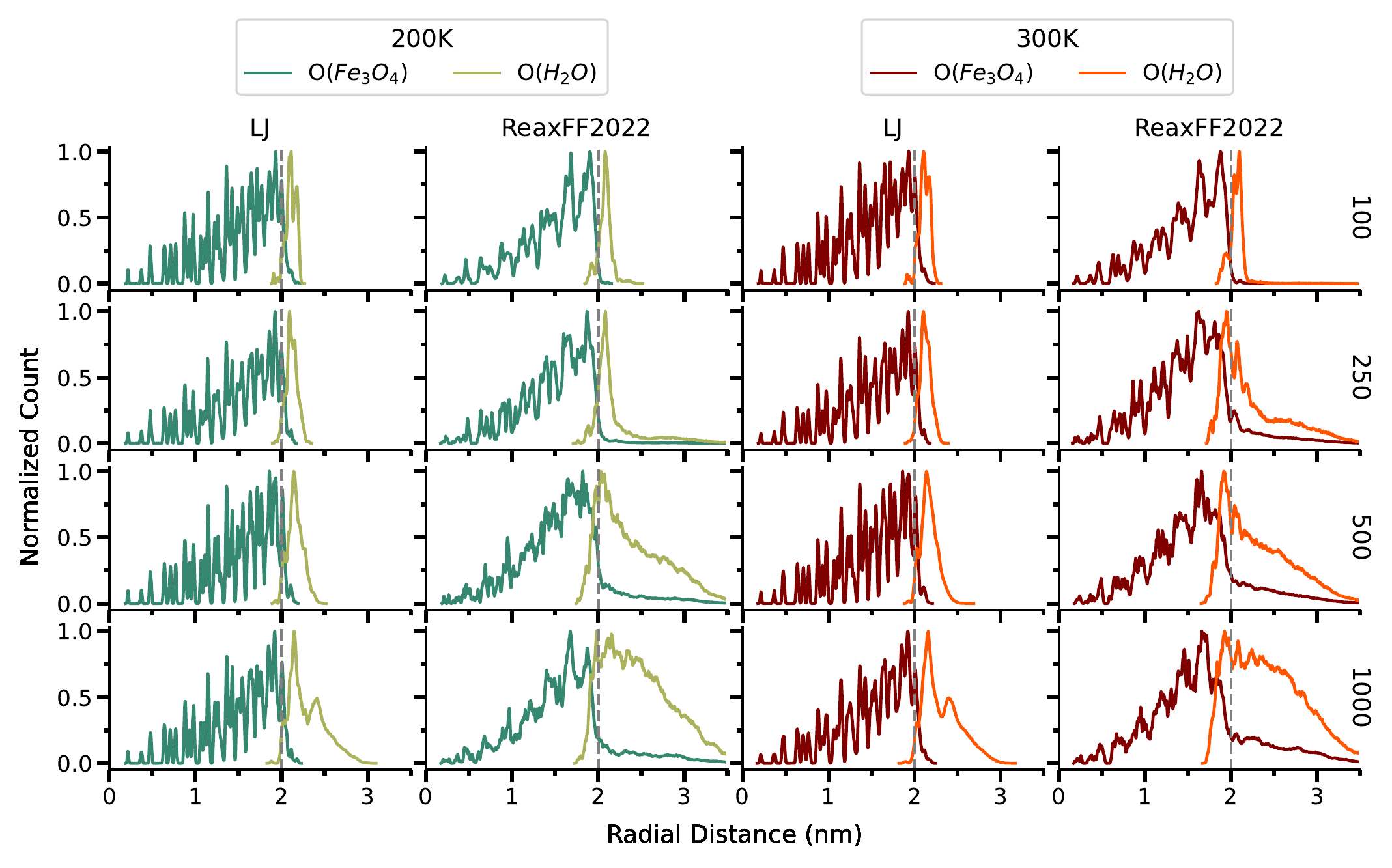}
\caption{Radial distribution histogram of \ce{Fe3O4} and \ce{H2O} oxygen with different amounts of water and temperature. Similar to Figure \ref{fig: 2nm_nanoparticle}, the results are from a $r=\qty{2}{nm}$ nanoparticle. Each row shows the nanoparticle with 100, 250, 500, and 1000 \ce{H2O} surface molecules. The first two columns are of a particle at \qty{200}{K} using the LJ and ReaxFF2022, respectively. The last two columns are the corresponding result for a system at \qty{300}{K}.} \label{fig: H2O_RDF}
\end{figure}

We calculated a radial distribution histogram that accounts for every atom's radial distance relative to the center of mass. Figure \ref{fig: H2O_RDF} shows the result at \qty{200}{K} and \qty{300}{K} for both force fields. The plot shows the normalized histogram for oxygen (see also Figure \ref{fig: H2O_RDF2} for the iron and hydrogen components). The result confirms that for the LJ force field, the \ce{H2O} molecules are attached to the nanoparticle with low diffusion. The nanoparticle keeps the same structure.

On the other hand, when using the ReaxFF2022 force field, the nanoparticle reacted to form a hydroxide, which caused the \ce{Fe3O4} to lose its original structure. Some \ce{H2O} molecules enter the crystal structure, and at the same time, \ce{H2O} molecules travel to the edge of the numerical domain. There are \ce{Fe3O4} oxygen atoms reaching the edge of the numerical domain as well. We consider the reaction between \ce{Fe3O4} and \ce{H2O} to be a simulation artifact. The reaction is exceedingly quick. The ReaxFF2022 force field was trained at higher temperatures and pressures than those observed in experiments to achieve an obvious corrosion layer in a short time. For this reason, the LJ force field appears to yield more consistent results.

\subsubsection{Temperature dependency}
\label{subsec: surface_water_temperature}

\begin{figure}[!ht]
\centering
    \includegraphics[width=0.85\textwidth]{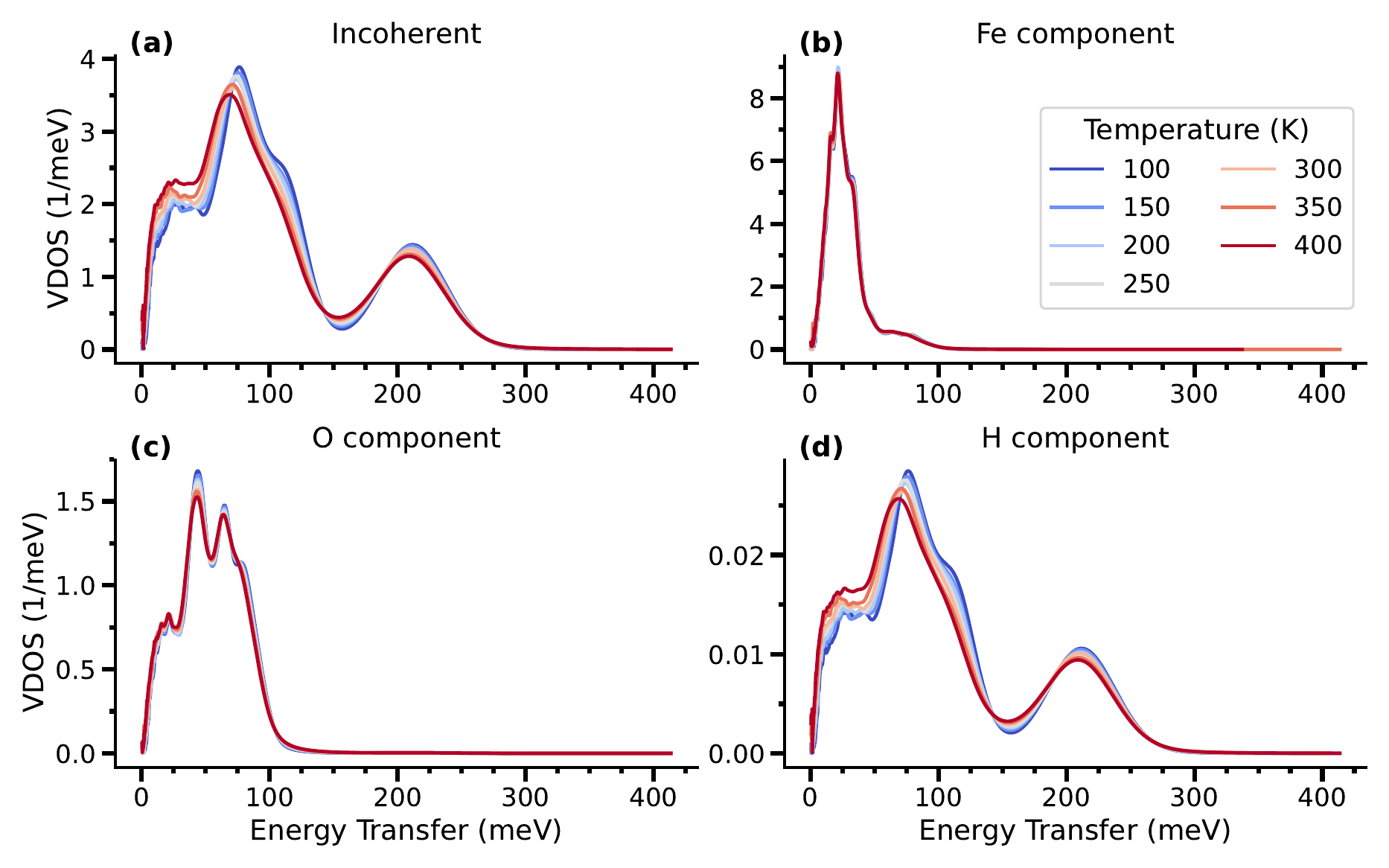}
\caption{Vibrational density of states temperature effect in nanoparticle with \ce{H2O}. The results are from a $r=\qty{2}{nm}$ particle with added 250 \ce{H2O} molecules. Panel \textbf{(a)} shows the incoherent VDOS at different temperatures. Panels \textbf{(b)} to \textbf{(d)} show the partial VDOS for \ce{Fe}, \ce{O} and \ce{H}, respectively.} \label{fig: H2O_temperature}
\end{figure}

We simulated a nanoparticle with 250 \ce{H2O} molecules and explored the effect of the temperature on the VDOS. Figure \ref{fig: H2O_temperature} shows the incoherent scattering and the iron, oxygen, and hydrogen components. Similarly to the result from section \ref{subsec: temperature_dependency}, the \ce{Fe} and \ce{O} VDOS are almost temperature-independent. However, the \ce{H} component shows a progressive red shift of the peak near \qty{75}{meV} when increasing the temperature. The peak near \qty{200}{meV} is present in the bulk simulations using the TIP4P2005f water model, but not in other non-flexible water models.

\section{Conclusions}
\label{sec: conclusions}

In this study, we simulated magnetite nanoparticles using molecular dynamics and characterized the resulting phonon density of states as a function of force field, nanoparticle size, and temperature. We examined a cluster of particles and the effects of adding \ce{H2O} on the nanoparticle surface. We compared the results with inelastic X-ray data, time-of-flight inelastic neutron scattering, and density functional theory phonon calculations. 

The tested force fields are good at reproducing some experimental phonon density of states features. However, none of the force fields are capable of fitting the overall density profile. Remarkably, the simple Lennard-Jones and Buckingham-based force fields are good at modeling the X-ray and neutron scattering results when calculating the incoherent and coherent phonon density of states, respectively. However, the iron component is shifted to higher energies relative to the experimental result. In contrast, the more sophisticated reaxff-based simulations show a very accurate iron peak, but an oxygen peak that is significantly shifted to lower energies. The Lennard-Jones-based force field correlates well with density functional theory phonon calculations. Overall, we consider it a useful model for magnetite nanoparticle simulations. 

The particle size has a significant impact on the resulting phonon density of states. The primary changes are attributed to surface-strain effects and an increased phonon scattering rate. The confinement produces a considerable broadening in the PDOS, which increases as particle size is decreased. A more pronounced slope at low energies represents Rayleigh surface phonons, which become more pronounced as particle size is decreased. Finally, the phonons slightly soften to lower energies as particle size is decreased. The softening and broadening become more apparent at higher energies. The nanoparticles themselves do not appear to change shape as size is decreased. 

We found that the PDOS is relatively temperature-independent in the \qty{100}{K} to \qty{400}{K} range. There is a slight softening of phonon modes, which becomes more pronounced in the higher energy oxygen region of the spectra. This is attributable to a low amount of phonon anharmonicity. The temperature should also have an impact on the magnetic properties of the particle. We left this as an open question since we did not simulate spin-lattice interactions.  

The isolated single particle is a good model of a cluster of particles. We did not observe a significant difference in the calculated PDOS between a single nanoparticle and a cluster of nanoparticles. 

The addition of \ce{H2O} molecules to the nanoparticle significantly affects the calculated PDOS. The TIP4P2005f water model, with a Lennard-Jones-based potential, produces a PDOS that changes progressively when \ce{H2O} molecules are added. The molecules are firmly attached to the nanoparticle surface. We observed low \ce{H2O} diffusion. In contrast, the reaxff-base simulation shows significant \ce{H2O} diffusion and a high rate of reactions between \ce{Fe3O4} and \ce{H2O}. Evaluating which of the force fields is the most accurate or realistic model would require further experimental results. 

Overall, we advance the understanding of magnetite nanoparticles by characterizing the PDOS under several common environmental factors. The studied system is complex and requires further investigation. In particular, the simulation of magnetic properties should influence the PDOS. Similarly, simulating a more realistic nanoparticle structure that includes a $\gamma$-\ce{Fe2O3} shell and incorporating other surface adsorbates can be useful for specific applications. We leave those topics for future work.

\section{Acknowledgment}
\label{sec: acknowledgment}

This research was undertaken with the assistance of resources and services from the National Computational Infrastructure (NCI), which is supported by the Australian Government. This work was supported by resources provided by the Pawsey Supercomputing Research Centre’s Setonix Supercomputer, with funding from the Australian Government and the Government of Western Australia. We acknowledge the support of the Australian Government in providing access to the Australian Centre for Neutron Scattering, which is partly funded through the National Collaborative Research Infrastructure Strategy (NCRIS).

\appendix
\section{Convergence test}
\label{app: convergence_test}

We performed a convergence test to determine the optimal timestep, simulation time, and sampling size. In each case, we followed the protocol outlined in Section \ref{sec: methods} with a single temperature value, $T_i=\qty{100}{K}$. We use a $r=\qty{2}{nm}$ particle. We use the CB1 force field (see Table \ref{tab: forcefields}).

\subsection{Timestep}
\label{app: convergence_timestep}

\begin{figure}[!ht]
\centering
    \includegraphics[width=\textwidth]{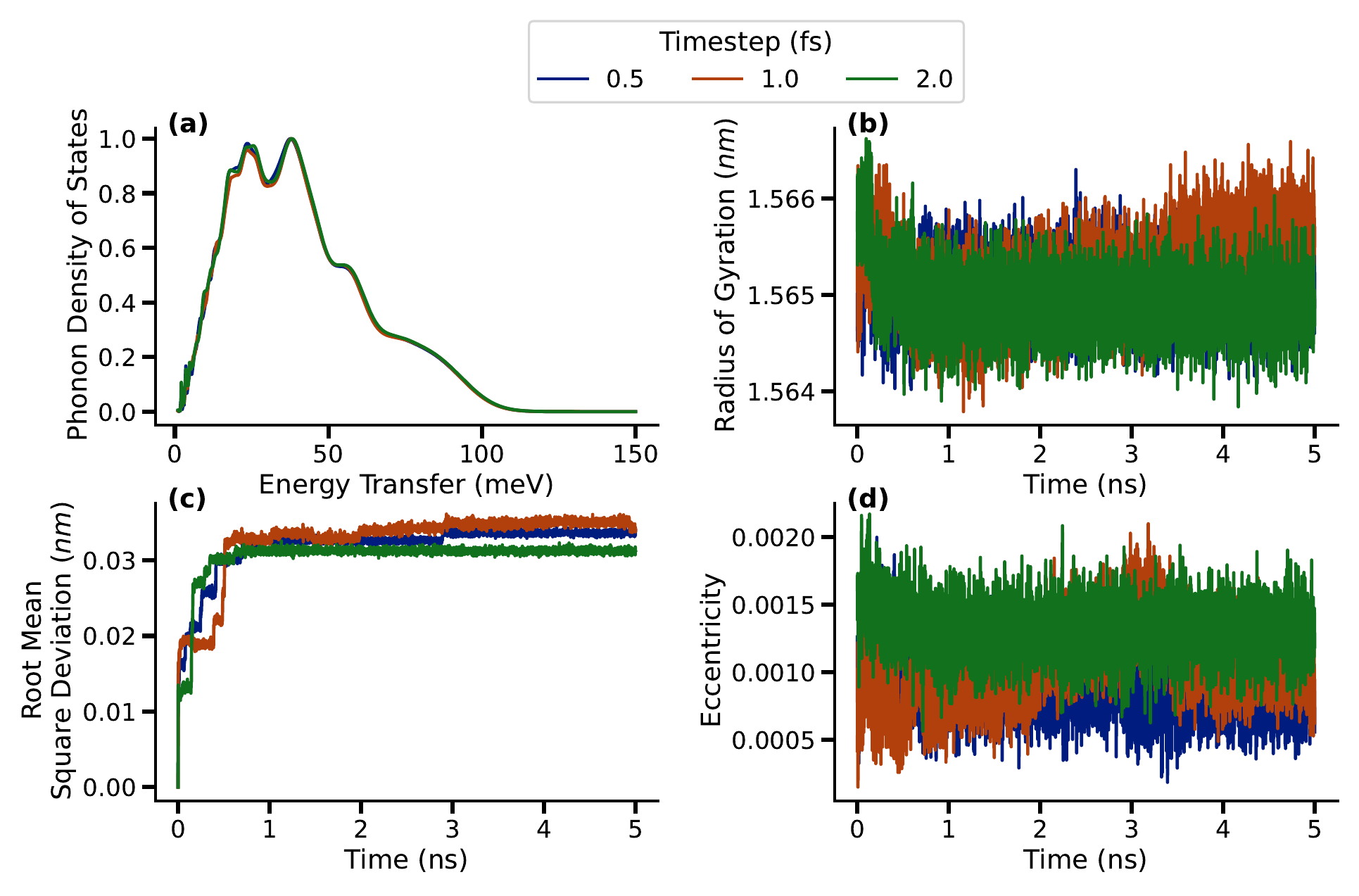}
\caption{Timestep convergence test. Panel \textbf{(a)} shows the normalised phonon density of states. Panel \textbf{(b)} shows the radius of gyration. Panel \textbf{(c)} shows the root mean-squared deviation. Panel \textbf{(d)} shows the eccentricity. } \label{fig: timestep_conv}
\end{figure}

We tested three values for the timestep: \qty{0.5}{fs}, \qty{1.0}{fs} and \qty{2.0}{fs}. Figure \ref{fig: timestep_conv} shows the results. The PDOS shows excellent convergence (Figure \ref{fig: timestep_conv}-\textbf{(a)}). There is no noticeable difference between the three results. The radius of gyration shows a minimal variability with similar mean and standard deviation values (see Figure \ref{fig: timestep_conv}-\textbf{(b)} and Table \ref{tab: rog_conv}). This shows that the particle was stable during the simulation for the three timesteps. The root mean-squared deviation shows large deviations relative to the initial position (see Figure \ref{fig: timestep_conv}-\textbf{(c)}. This is due to a rotation of the particle. We started the simulation with a zero linear and angular momentum velocity distribution. However, thermal fluctuations introduce random rotation of the particle. Nonetheless, the root mean-squared deviation range for the three simulations is consistent. The eccentricity is close to zero for the three simulations (see Figure \ref{fig: timestep_conv}-\textbf{(d)} and Table \ref{tab: rog_conv}). This shows that the particle keeps a spherical shape. Based on these results, we selected a time step of \qty{1}{fs} for all simulations. 

\begin{table}[h]
    \centering
    \begin{tabular}{ccc|cc}
        & \multicolumn{2}{c}{Radius of gyration} & \multicolumn{2}{c}{Eccentricity} \\
         Timestep (\unit{fs})& $\mu_{rog}(\unit{nm})$ & $\sigma_{rog}(\unit{nm})$ &$\mu_{ecc}$ & $\sigma_{ecc}$\\
         \hline \hline
        0.5 & 1.5651 & 0.0003 & 0.0009 & 0.0002 \\
        1.0 & 1.5653 & 0.0004 & 0.0010 & 0.0003 \\
        2.0 & 1.5649 & 0.0004 & 0.0013 & 0.0002 \\
                 
\end{tabular}
    \caption{Radius of gyration and eccentricity mean and standard deviation.}
    \label{tab: rog_conv}
\end{table}

\subsection{Simulation time}
\label{app: convergence_simulation_time}

\begin{figure}[!ht]
\centering
    \includegraphics[width=\textwidth]{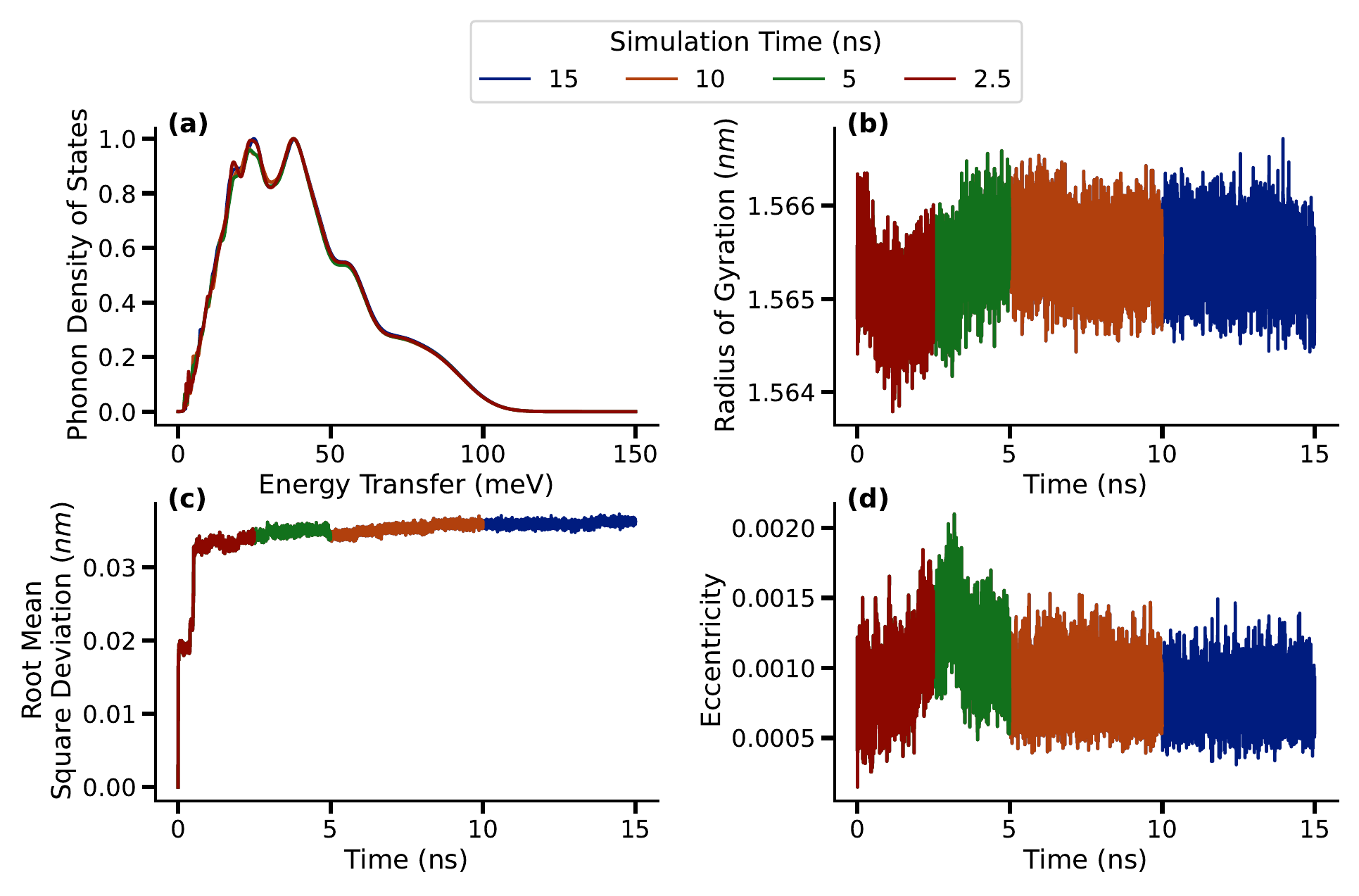}
\caption{Simulation time test. Panel \textbf{(a)} shows the normalised phonon density of states. Panel \textbf{(b)} shows the radius of gyration. Panel \textbf{(c)} shows the root mean-squared deviation. Panel \textbf{(d)} shows the eccentricity. } \label{fig: simulation_time}
\end{figure}

To assess the optimal simulation time, we performed three simulations that lasted \qty{2.5}{ns}, \qty{5}{ns}, \qty{10}{ns}, and \qty{20}{ns}. This test aims to see if there is a significant change in the phonon density of states calculated at different simulation times and to evaluate any possible instability in longer simulations. The simulation time does not significantly impact the phonon density of state (see Figure \ref{fig: simulation_time}-(\textbf{a})). The radius of gyration and eccentricity are relatively stable (see Figure \ref{fig: simulation_time}-(\textbf{b})\&(\textbf{d})). The root mean square deviation shows a slow drift due to the mentioned slow rotation (see Figure \ref{fig: simulation_time}-(\textbf{c})). We expect any stability or structural change to show relatively quickly in the simulation. Therefore, we decided to perform the production runs up to \qty{2.5}{ns} for computation economy. Since the phonon density of states does not show significant dependence on the sampling time, we performed the phonon sampling before the production runs. 

\subsection{Sampling frequency}
\label{app: sampling_frequency}
\begin{figure}[!ht]
\centering
    \includegraphics[width=\textwidth]{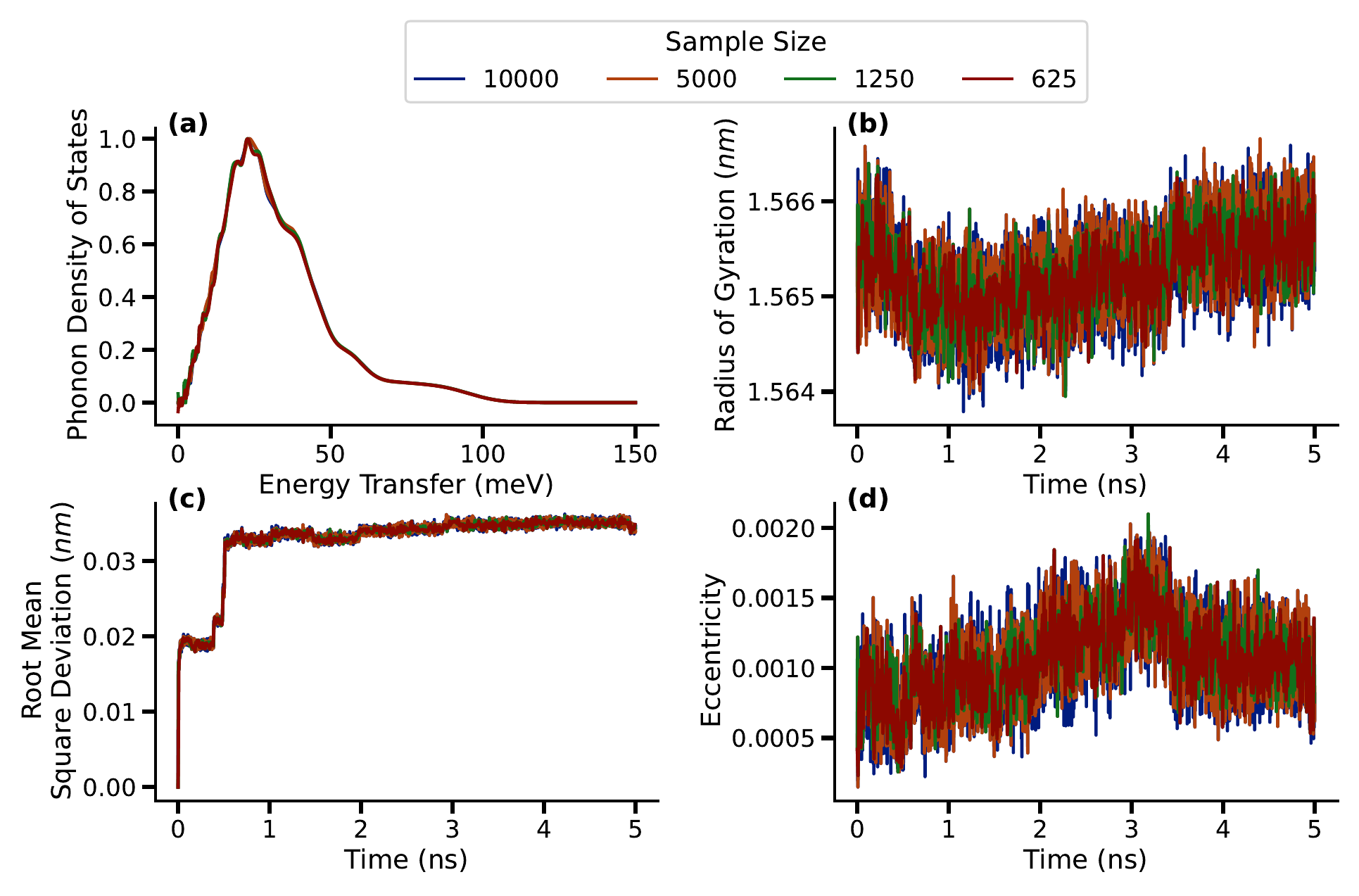}
\caption{Sampling convergence test. Panel \textbf{(a)} shows the normalised phonon density of states. Panel \textbf{(b)} shows the radius of gyration. Panel \textbf{(c)} shows the root mean-squared deviation. Panel \textbf{(d)} shows the eccentricity. } \label{fig: sampling_frequency}
\end{figure}

This test aims to determine the optimal sampling frequency. We tested three sampling sizes for stability and phonon calculation runs: 2500, 5000, and 10,000 sample points. Figure \ref{fig: sampling_frequency} shows the result. Since the sampling size has a negligible impact on the result, we decided to use 5000 points for phonon calculations and 625 for production runs.

\section{Force fields parameters}
\label{app: force_field_parameters}
Here we reproduce the force field parameters from \cite{ff_LJ, ff_CB1, ff_RFF1, ReaxFF2022}. 
\begin{table}[h]
    \centering
    \begin{tabular}{cccc}
     Atom type & $\epsilon_{ii}(\unit{kcal.mol^{-1}})$ & $\sigma_{ii}(\unit{\angstrom})$ & $q(\unit{e})$ \\
     \hline
     \hline
      \ce{Fe2+}  & $9.0298\times 10^{-5}$ & 4.90620 &  1.050\\
      \ce{Fe3+}  & $9.0298\times 10^{-5}$ & 4.90620 &  1.575\\
      \ce{O2-}  & 0.1554 & 3.16554 & -1.050\\
    \end{tabular}
    \caption{Lennard-Jones parameters from \cite{ff_LJ}. The mixing rule is Lorentz-Berthelot: $\epsilon_{ij}=\sqrt{\epsilon_i\epsilon_j},\sigma_{ij}=\frac{1}{2}(\sigma_i+\sigma_j)$}
    \label{tab: LJ_parameters}
\end{table}

\begin{table}[h]
    \centering
    \begin{tabular}{cccc}
     Interaction & $A(\unit{eV})$ & $B(\unit{\angstrom})$ & $C(\unit{eV.\angstrom^6})$\\
     \hline
     \hline
      \ce{Fe2+}:\ce{O2-}  & 1515.42 & 0.2756 &  0 \\
      \ce{Fe3+}:\ce{O2-}  &  895.56 & 0.3099 &  0 \\
      \ce{O2-}:\ce{O2-}   & 7322.63 & 0.2301 &  38.532 \\
    \end{tabular}
    \caption{Buckingham parameters from \cite{ff_CB1}. The charges are $q(\unit{e})$=1.530, 2.295 and -1.530(\unit{e}) for \ce{Fe2+}, \ce{Fe3+} and \ce{O2-} respectively.}
    \label{tab: CB_parameters}
\end{table}

The ReaxFF parameters in LAMMPS format can be downloaded from \cite{GalavizGit2025}.
\section{Bulk water simulation}
\label{app: bulk_water_simulation}

\begin{figure}[!ht]
\centering
    \includegraphics[width=0.7\textwidth]{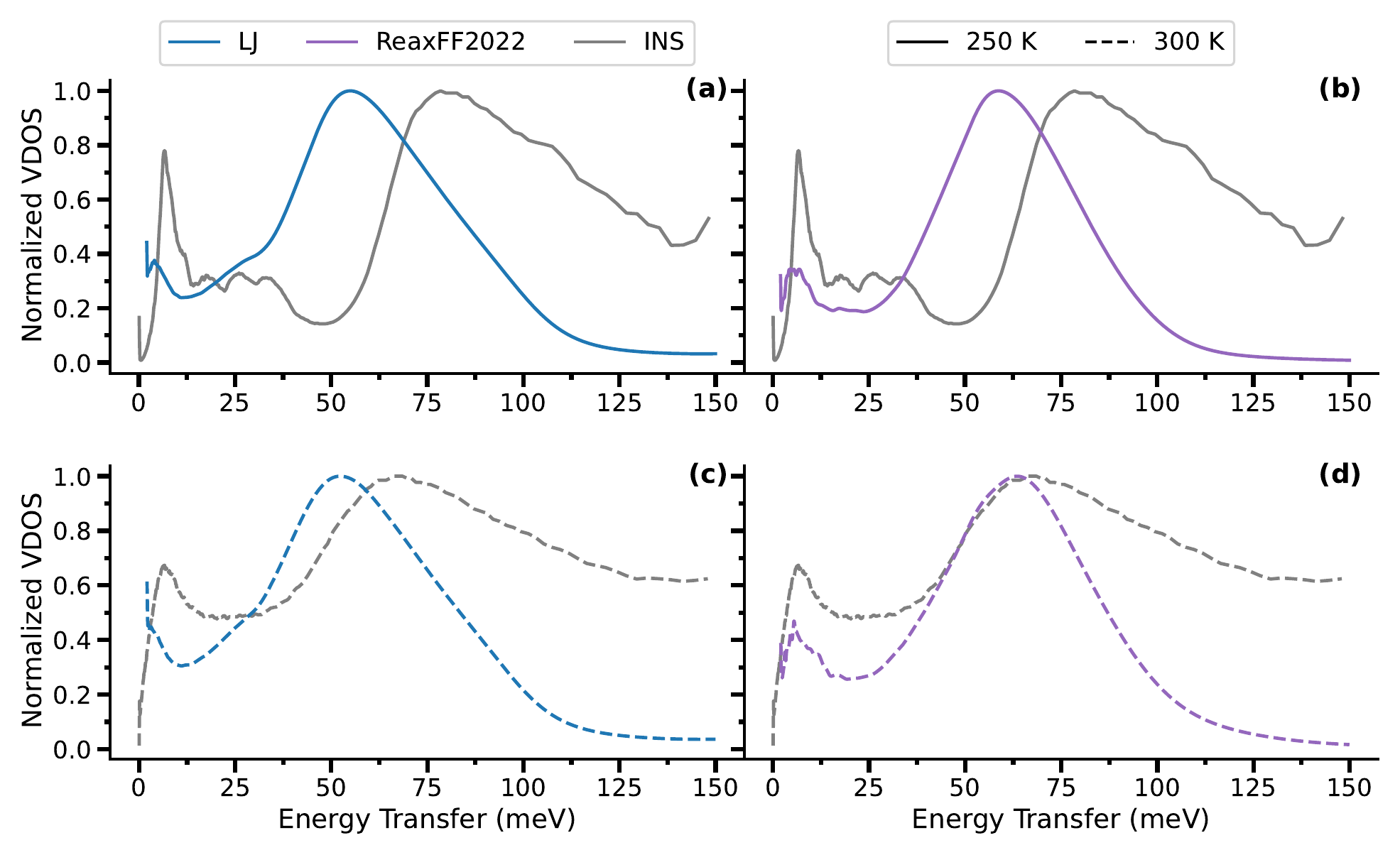}
\caption{VDOS of ice and water comparison between simulation and experiment.
In each panel, the grey line is the INS experimental result. Panel \textbf{(a)} shows the normalised PDOS of water at \qty{200}{K} for a simulation using the LJ force field. Panel \textbf{(b)} shows the result for the ReaxFF2022 force field. Panels \textbf{(c)} and \textbf{(d)} show the corresponding PDOS at \qty{300}{K}. } \label{fig: bulk_H2O_dos}
\end{figure}

Here, we examined bulk water phonon calculations using the TIP4P2005f and ReaxFF2005 force fields. The TIP4P2005f simulation was performed in GROMACS using a numerical cubic domain of side \qty{3.4}{nm} containing 1261 \ce{H2O} molecules (\qty{0.983}{kg.L^{-1}}). The simulation employs the protocol outlined in Section \ref{sec: methods}, with the exception that heating was performed using an NPT ensemble at \qty{1}{bar} for \qty{200}{ps}. The ReaxFF2022 simulation used a smaller cubic domain of size \qty{1.2}{nm} with 93 \ce{H2O} molecules (\qty{1.024}{kg.L^{-1}}). We compare these results with experimental TOF-INS (see \cite{JinFanSta2024} for the details). Figure \ref{fig: bulk_H2O_dos} shows the results. Both models exhibit poor performance in reproducing the low temperature; the experimental peak at around \qty{80}{meV} is shifted to lower energies by approximately \qty{20}{meV} to \qty{25}{meV}. The modeling of energies lower than \qty{25}{meV} is also inaccurate. For \qty{300}{K}, the results are better. The ReaxFF2022 accurately reproduces the peak positions. The low energy is still inaccurate in the LJ force field, in addition to a small softening in the \qty{60}{meV} peak. Overall, the tested water models are not accurate but are phenomenologically useful. 

\subsection{Additional water simulation results}
\label{app: additional_water_simulation_results}

\begin{figure}[!ht]
\centering
    \includegraphics[width=0.85\textwidth]{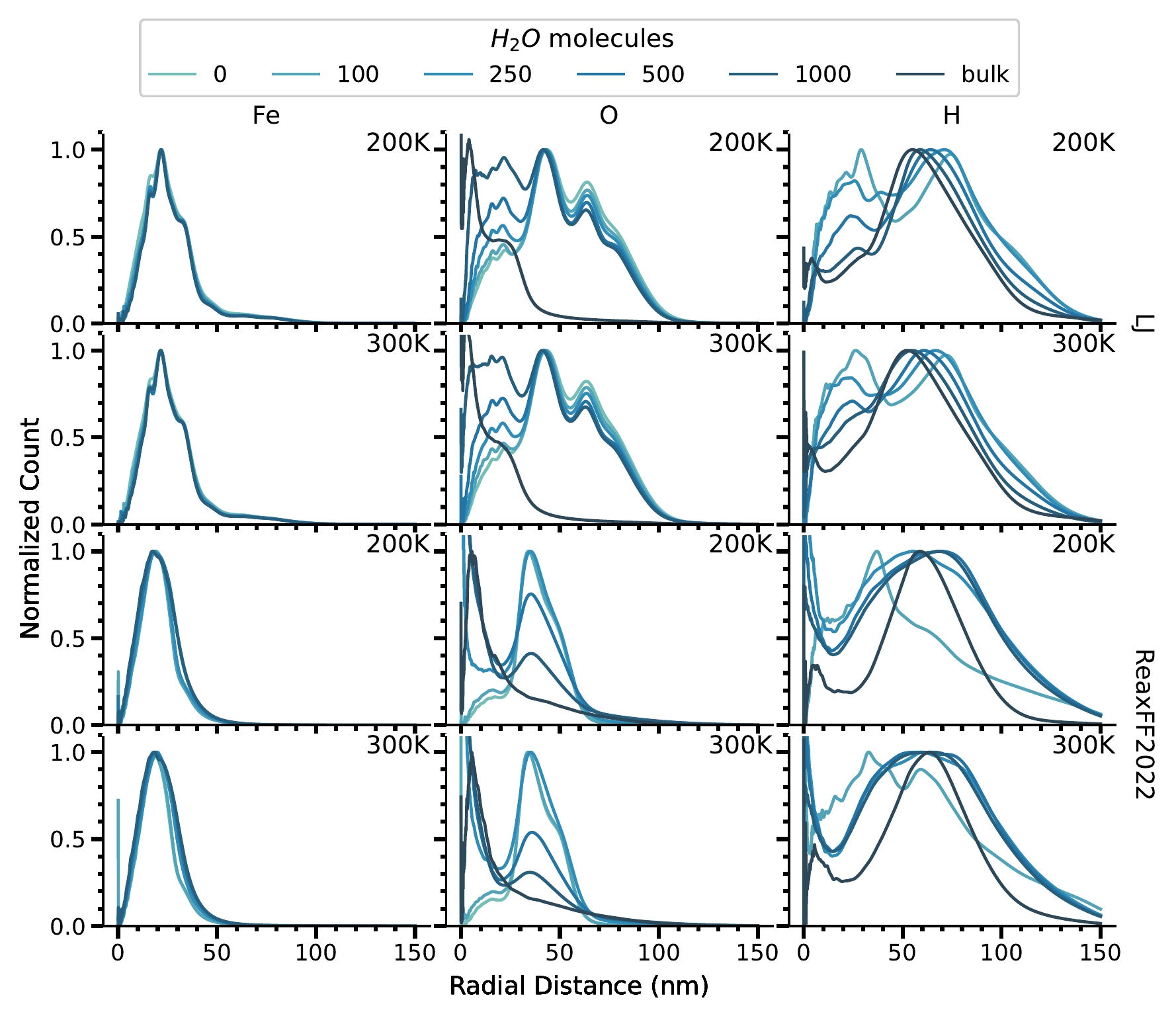}
\caption{Normalized PDOS for a particle with 0, 100, 250, 500, and 1000 \ce{H2O} surface molecules and bulk. The results are from a $r=\qty{2}{nm}$ nanoparticle. Each column shows the PDOS for the \ce{Fe}, \ce{O}, and \ce{H} components, respectively. The first two rows show the LJ simulation at \qty{200}{K} and \qty{300}{K}. The last two rows show the corresponding result for the ReaxFF2022 force field. } \label{fig: H2O_comp_dos}
\end{figure}

Figure \ref{fig: H2O_comp_dos} shows the partial PDOS. For both force fields, the \ce{Fe} component is independent of the amount of surface water. The interior of the particle is unaffected. On the other hand, in the LJ result, the \ce{O} component transition form modes at energies larger than \qty{25}{meV} to lower energies. This is primarily a surface effect, since in this simulation, the \ce{H2O} remains mostly as a coating on the particle. On the other hand, for the ReaxFF2022, there is a reaction between the \ce{H2O} and the \ce{Fe3O4}, and water diffusion. The oxygen peak at around \qty{40}{meV} diminishes when adding more particles.

\begin{figure}[!ht]
\centering
    \includegraphics[width=0.85\textwidth]{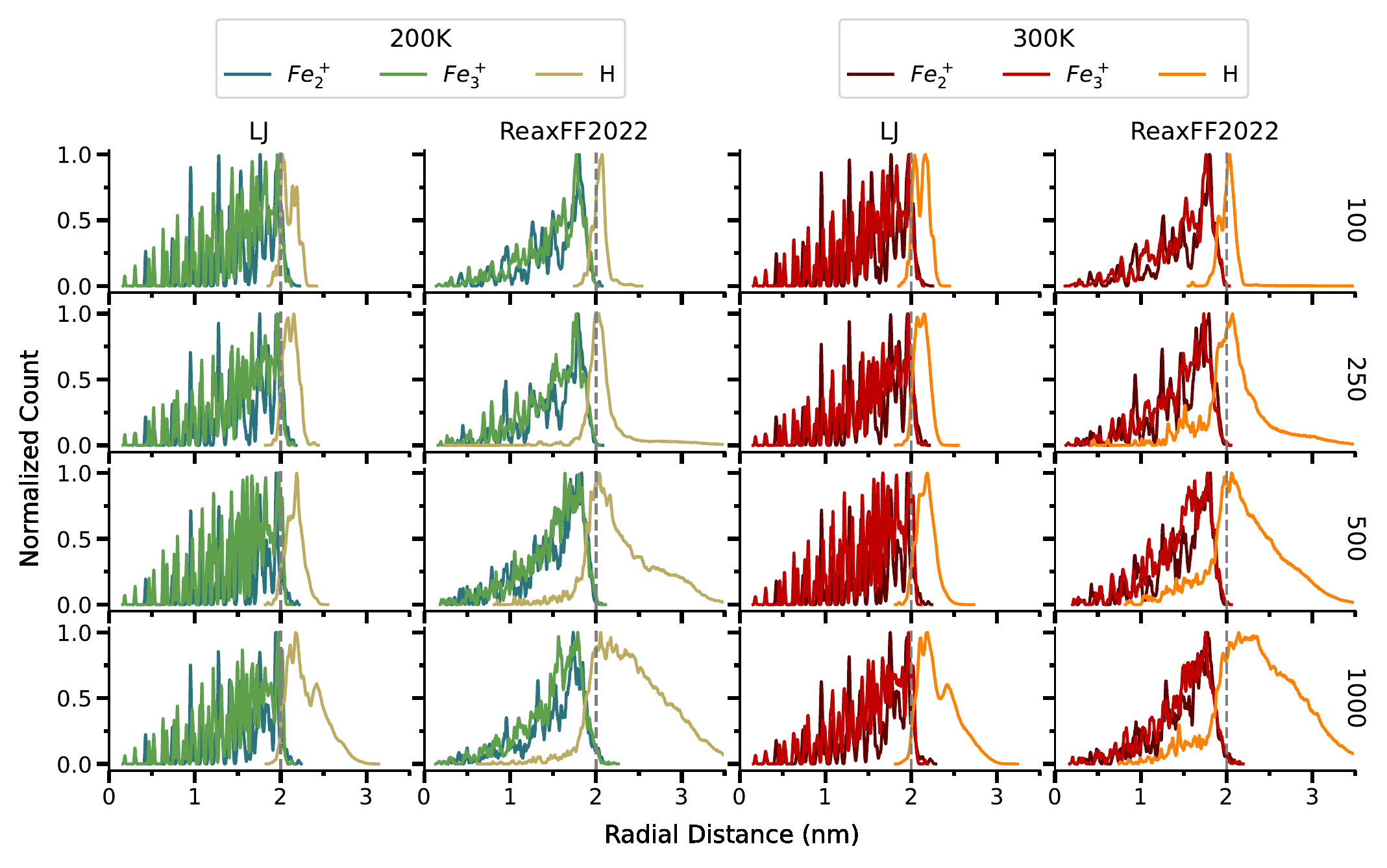}
\caption{Radial distribution histogram of \ce{Fe3O4} iron and \ce{H2O} hydrogen with different amounts of water and temperature. Similar to Figure \ref{fig: H2O_RDF}, the results are from a $r=\qty{2}{nm}$ nanoparticle. Each row shows the nanoparticle with 100, 250, 500, and 1000 \ce{H2O} surface molecules. The first two columns are of a particle at \qty{200}{K} using the LJ and ReaxFF2022, respectively. The last two columns are the corresponding result for a system at \qty{300}{K}.} \label{fig: H2O_RDF2}
\end{figure}

Figure \ref{fig: H2O_RDF2} shows the radial distribution histogram for iron and hydrogen. The result confirms that the iron distribution is not affected by the \ce{H2O} molecules. However, in the case of the ReaxFF2022 force field, we can see hydrogen reaching the nanoparticle's core as part of the occurring reactions. 

\bibliographystyle{elsarticle-num} 
\bibliography{main}







\end{document}